\documentclass[9pt,conference]{IEEEtran}
\IEEEoverridecommandlockouts

\usepackage{cite}
\usepackage{algorithm}
\usepackage{amsmath,amssymb,amsfonts}
\usepackage{bm}
\usepackage{csvsimple}
\usepackage{subcaption}
\usepackage{wasysym}
\usepackage[capitalise]{cleveref}
\usepackage{algorithmic}
\usepackage{graphicx}
\usepackage{diagbox}
\usepackage{tablefootnote}
\usepackage{textcomp}
\usepackage{textgreek}
\usepackage{comment}
\usepackage{booktabs}
\usepackage{xcolor}
\usepackage{todonotes}
\usepackage{soul}
\usepackage{url}
\def\BibTeX{{\rm B\kern-.05em{\sc i\kern-.025em b}\kern-.08em
    T\kern-.1667em\lower.7ex\hbox{E}\kern-.125emX}}
\begin{document}

\title{Computational Facilitation of Large Scale\\ Microfluidic Fuel Cell Architectures\\
\thanks{}
}


\author{\IEEEauthorblockN{Michel Takken\IEEEauthorrefmark{1} \qquad \qquad \qquad 
Robert Wille\IEEEauthorrefmark{1}\IEEEauthorrefmark{2}}
\IEEEauthorblockA{\IEEEauthorrefmark{1}Technical University of Munich (TUM), Arcisstra{\ss}e 21, 80333 M\"unchen, Germany\\
\IEEEauthorrefmark{2}Munich Quantum Software Company GmbH, M\"unchener Str. 34, 85748 Garching bei M\"unchen, Germany\\
michel.takken@tum.de \qquad \quad \qquad robert.wille@tum.de\\
\url{https://www.cda.cit.tum.de/research/microfluidics/}}}

\maketitle

\begin{abstract}
Hydrogen fuel cells are a key technology in the transition toward carbon-neutral energy systems, offering clean power with water as the only byproduct. Microfluidic fuel cells, which operate at the microliter scale, are an emerging variant that offer fine control over fluid and thermal dynamics, along with compact, efficient designs. However, scaling these systems to meet practical power demands remains a major challenge—particularly due to the limitations of conventional simulation methods like \emph{Computational Fluid Dynamics} (CFD), which are computationally expensive and scale poorly. In this work, we propose a reduced-order simulation method that models the behavior of individual microfluidic fuel cells and efficiently extends it to large scale stacks. This approach significantly reduces simulation time while maintaining close agreement with detailed CFD results. The method is validated, evaluated for scalability, and discussed in the context of ongoing advancements in microfluidic fuel cell fabrication. The obtained results demonstrate that this abstraction can support the design and development of scalable microfluidic fuel cell systems and, for the first time, the consideration of first macroscale instances of practical value.
\end{abstract}


\section{Introduction}
\label{sec:Introduction}
\begin{figure}[t]
    \includegraphics[width=0.95\linewidth]{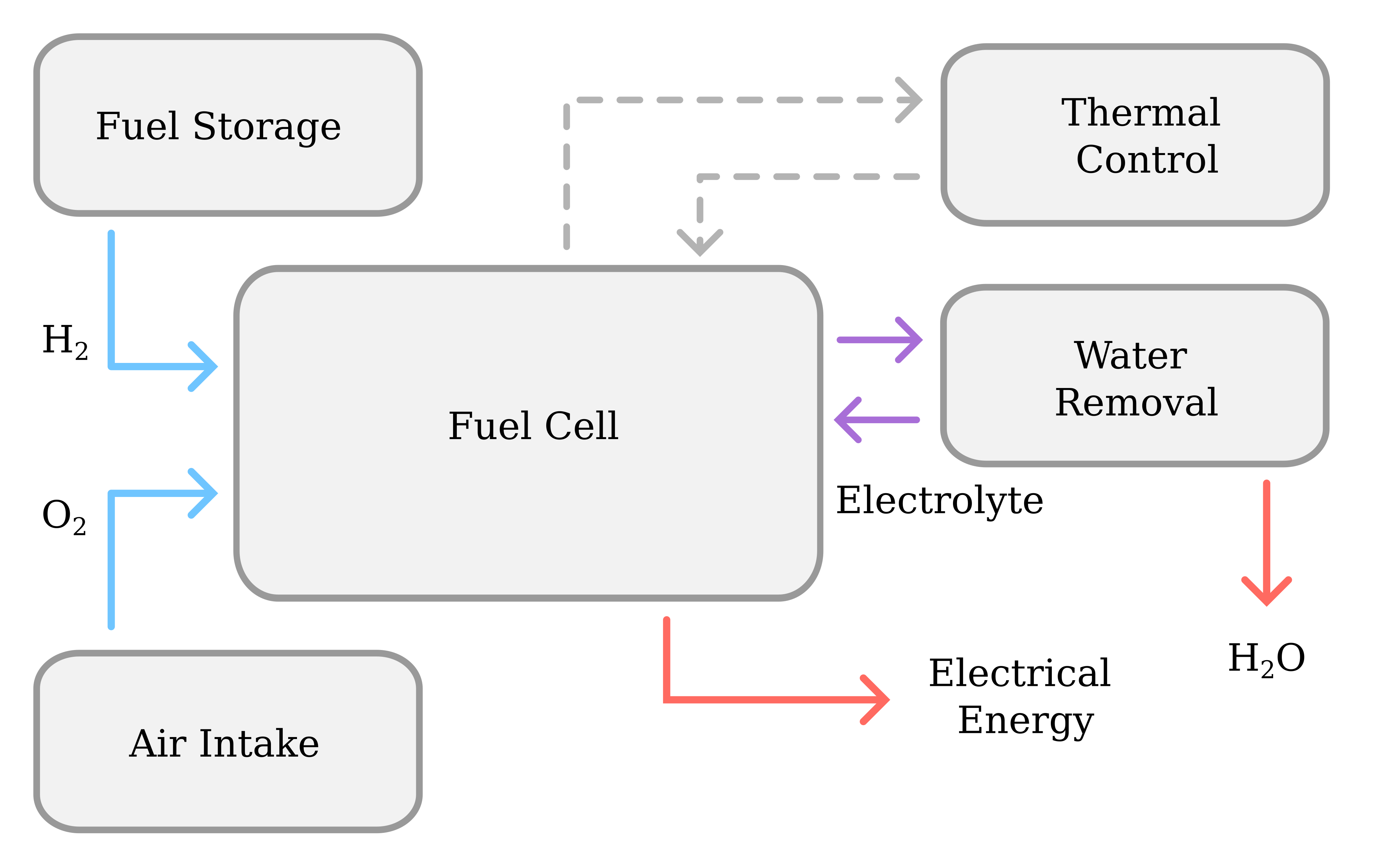}
    \caption{A schematic representation of a conventional air-breathing fuel cell system. Besides the fuel cell, the system is composed of various subsystems, including: A fuel and air intake subsystem, a water removal subsystem, and thermal control subsystem.}
    \label{fig:FuelCellSystem}
\end{figure}

In order to reduce the risks and impacts of climate change~\cite{ParisAgreement}, many economies have set carbon neutrality goals and developed strategies to achieve those~\cite{EUGreenDeal,JapanCarbonNeutral,ChinaCarbonNeutral,CaliforniaB55}. 
Most of which have identified hydrogen and fuel cells as key energy technologies of the 21\textsuperscript{st}~century~\mbox{\cite{fan2021recent,kovavc2021hydrogenperspective,capurso2022hydrogenperspective}}. Fuel cells are electrochemical cells that convert chemical reactants into electrical energy and byproducts. A schematic illustration of a hydrogen fuel cell system is provided in \cref{fig:FuelCellSystem}, which shows the inputs to the system (i.e., the hydrogen and oxygen) in blue, and the outputs (i.e., electrical energy and water) in red. During operation, the fuel cell produces heat, which can optionally be managed using, e.g., a heat sink or a coolant combined with a thermal control subsystem. Furthermore, a hydrogen fuel cell generally contains an electrolyte that is necessary for the electrochemical reactions. The products of that reaction must be removed from the system, which is illustrated by the water removal subsystem. 

The advantages of the system are evident, as it produces electrical energy with water as its only byproduct.
Unlike batteries, fuel cells do not require charging, but rather a replenishment of the chemical reactants, i.e., fuel. The use of fuel cells is versatile, ranging from mobile applications, such as aviation~\cite{yusaf2024sustainable}, shipping~\cite{atilhan2021green}, and \emph{Unmanned Aerial Vehicles}~(UAVs,~\cite{shen2024review}), to stationary applications, such as domestic power generation and \emph{Combined Heat and Power}~(CHP) production on residential blocks~\cite{aminudin2023overview,wilberforce2016advances}. 
Common fuel cell types for these applications are the \emph{Proton Exchange Membrane Fuel Cell}~\mbox{(PEMFC, \cite{mancino2023pem})} or the \emph{Solid Oxide Fuel Cell}~\mbox{(SOFC, \cite{singh2021solid})}.
However, before a large scale adoption of fuel cells is possible, many technological challenges still need to be overcome. These include challenges with flow and/or thermal control, flooding, or membranes that degrade over time~\cite{Ohayre2016}. 

In parallel, microfluidic devices (i.e., devices that operate fluids in the nano-/microliter scale) are being developed and have succesfully been employed for \mbox{\emph{Point-of-Care}} (PoC,~\cite{yang2022microfluidic}), \mbox{\emph{Lab-on-Chip}}~(LoC,~\cite{sharma2022microfluidics}), and \mbox{\emph{Organ-on-Chip}}~(OoC,~\cite{nahak2022advances}) applications. The capability of microfluidic devices to manipulate fluids on a small scale allows for a better control of, e.g., flow and temperature, making them a promising platform for (membraneless) fuel cells. In fact, microfluidic fuel cells are actively being developed~\mbox{\cite{wang2023two,li2024narrow,luo2023breakthrough,ouyang2023performance,liu2024high,han2025stacked,brushett2010investigation,hollinger2013manufacturing,armenta2016improved,shaegh2012air,yang2019flexible,martins2018bleaching,li2023counter}} 
since their introduction \cite{ferrigno2002membraneless}. Similar to conventional fuel cells, they are successfully being stacked into microfluidic fuel cell stacks, with up to 6 cells, to provide more power \cite{wang2016high,kumar2024microfluidic,zhang2024investigation,dai2025combined,ho2013planar,ibrahim2016situ,moreno2017evolution}. In addition, recent developments show efforts made to integrate micro-scale pumps on \emph{Micro-Electromechanical Systems} (MEMS)~\cite{seidl2024fully}, purify water in electrochemical microfluidic devices~\cite{park2025energy}, and manage heat in the scale-up of microfluidic devices~\cite{sarbaland2025temperature}. These form micro-scale equivalents of the supporting sub-systems (shown in \cref{fig:FuelCellSystem}) that are necessary to operate fuel cells efficiently.
Overall, these developments indicate that microfluidic fuel cells are an emerging technology that show great potential and may play a role in achieving carbon neutrality.


However,  in order to meet the power production of their macrofluidic counterparts, a major challenge that still remains is scaling. Thus far, this was mainly a fabrication issue, but with the advances in micro-fabrication methods~\cite{scott2021fabrication}, (rapid) 3-D printing techniques for microfluidic devices~\mbox{\cite{su20233d,collingwood2023high,gonzalez2022current,nielsen20203d}} 
and the advances in integrating electrodes in microfluidic devices~\mbox{\cite{selemani20243d,costa2022stereolithography,duarte20173d}}, substantial steps towards the realization of large scale microfluidic devices have been made. As a result, scaling fuel cell stacks has progressively become a design concern. To provide an intuition of the required scaling in practical purposes, \cref{tab:SystemPeakPower} lists some examples of required stack sizes needed for different systems, considering a representative cell size of \mbox{0.25x2 mm} and a peak power density of 100 mW/cm\(^2\)~\cite{Wang2021FC}. It is obvious that, with stack sizes like that, the realization of microfluidic fuel cell stacks requires significant design capabilities. 

Simulation can significantly aid in the design of corresponding devices \cite{Grimmer2018} and, in this regard, are key as they allow to explore different designs and check/validate them prior to fabrication. Unfortunately, state-of-the-art simulation methods for microfluidic fuel cells~\cite{andaluz2023modeling,ouyang2023performance,wu2024microfluidic}, such as those using \emph{Computational Fluid Dynamics}~\mbox{(CFD, \cite{Ferziger2019})}, are time-consuming and scale poorly~\mbox{\cite{Takken2022,Wu2024}}. This makes it practically impossible to simulate large scale fuel cell stacks.

In an effort to address the current situation, this work proposes an alternate simulation approach that utilizes a higher level of abstraction than the state of the art and, by this, allows to efficiently simulate large scale microfluidic fuel cell stacks, using a \mbox{reduced-order} model while maintaining the desired accuracy. In the remainder of this work, 
we carefully review the necessary background and governing equations for a representative microfluidic fuel cell, in \cref{sec:Background}. Afterwards, the abstract simulation method is explained in a \mbox{two-fold} fashion: first the abstraction is explained for a single cell in \cref{sec:SingleCell}, based on which the abstraction is extended to a stack in \cref{sec:Stack}. Subsequently, the abstraction is validated against simulation results from CFD in \cref{sec:Justification}, followed by a scaling evaluation in \cref{sec:Performance}. Finally, the work is concluded in \cref{sec:Conclusion}.

\begin{table}[t]
    \small
	\centering
	\caption{Power requirement and stack sizes of\\ different systems.}
	\label{tab:SystemPeakPower}
	\def\arraystretch{1.0}
    \setlength{\tabcolsep}{2pt}
    \begin{tabular}{@{\,}l@{\quad}rrlr}
        \textit{\textbf{System}} & \multicolumn{3}{c}{\shortstack{Power \\ Requirement {[kW]}}} & \shortstack{Stack Size\\ {[No. cells]}} \\
        \midrule
        UAV & \hspace{8mm}
        & 0.5 & \cite{gong2023modeling} & 1.0\(\times \text{10}^\text{6}\)
        \\
        US Household* &
        & 3.0 & \cite{UsHousehold} & 6.0\(\times \text{10}^\text{6}\)
        \\
        Electric Car &
        & 210.0 & \cite{BMWi4} & 4.2\(\times \text{10}^\text{8}\)
        \\
        Subregional Aircraft &
        & 2000.0 & \cite{tom2021commercial} & 4.0\(\times \text{10}^\text{9}\)
        \\
        \bottomrule
    \multicolumn{5}{l}{* \footnotesize Average power requirement.}
    \end{tabular}
\end{table}

\vspace{5mm}
\section{Microfluidic Fuel Cell Model:\\ Governing Equations}
\label{sec:Background}
The microfluidic fuel cell is modeled through a set of governing equations for electrochemical reactions and fluid dynamics that describe the system. To simplify the discussion, we consider a \mbox{single-flow} \emph{Alkaline Fuel Cell}~\mbox{(AFC,  \cite{jayashree2007microfluidic,brushett2010carbon,naughton2011carbonate,wang2016toward})} 
as a representative. The AFC consists of an alkaline electrolyte with the hydroxide ion~(OH\(^\text{--}\)) as charge carrier. The \mbox{half-reactions} of the electrochemical reaction are given by the \emph{Hydrogen Oxidation Reaction}~(HOR) at the anode and the \emph{Oxygen Reduction Reaction}~(ORR) at the cathode:
\begin{align}
    \label{eq:HOR} \text{\text{Anode HOR:}} \quad \qquad \;\; H_2 + 2OH^- &\xrightarrow{} 2H_2O + 2e^-\\
    \label{eq:ORR} \text{\text{Cathode ORR:}} \quad O_2 + 2H_2O + 4e^- &\xrightarrow{} 4 OH^-.
\end{align}
\noindent The reactants of the full electrochemical reaction are oxygen and hydrogen and the product is water. This reaction will be used in the remainder of this work to illustrate the modeling and simulation approach for a microfluidic fuel cell. In this section, the governing equations will be given for the electrochemical reactions and fluid dynamics, respectively. Afterwards, the resulting polarization curve is discussed.

\subsection{Electrochemical Reactions}

The operating state of a microfluidic fuel cell is defined by a combination of operating voltage and current density, which are coupled. The equilibrium potential for a cell describes the potential in steady-state operation and is given by 
\begin{equation}
    \label{eq:cellPotential}
    E^\text{cell} = E_0^C - E_0^A - \eta^C - \eta^A.
\end{equation}
\noindent We define~\(S = \{A,C\}\), where~\(A\) and~\(C\) denote the anode and cathode electrode, respectively, and we define~\(e\) as an element of~\(S\). The equilibrium potential for an electrode is given by \(E_0^e\), and~\(\eta^e\) denotes the corresponding electrode's overpotential.

The equilibrium potential for an electrode is given by the Nernst equation \cite{Bard2022,Ohayre2016,Wu2024}, as
\begin{equation}
\label{eq:Nernst}
    E_0^e = E_0^{e,\text{ref}} - \frac{RT}{nF}ln\left( \displaystyle \prod_{i \in \text{Red.}} \left(\frac{c_i^e}{c_i^\text{ref}}\right)^{\theta_i} \displaystyle \prod_{i \in \text{Ox.}} \left(\frac{c_i^e}{c_i^\text{ref}}\right)^{-\theta_i}\right),
\end{equation}
\noindent where \(E_0^{e,\text{ref}}\) is the electrode's reference equilibrium potential, \(R\) and \(F\) are the gas constant and Faraday's constant, respectively, \(n\) is the number of electrons transferred in the reaction, and \(T\) is the temperature. The local concentration on the electrode surface and the reference concentration are denoted by~\(c_i^e\) and~\(c_i^\text{ref}\), respectively, and~\(\theta_i\) is the stoichiometric number of species~\(i\). The species~\(i\) either undergoes \emph{Reduction} (Red.) or \emph{Oxidation} (Ox.) at the electrode. 

The overpotentials are related to the operating current density through the \mbox{Butler-Volmer} equation \cite{Bard2022,Ohayre2016,Wu2024}, which is given for the electrodes as

\begin{multline}
\label{eq:ButlerVolmer}
    j = j_0^e \Bigg[ \displaystyle \prod_{i \in \text{Red.}} \left( \frac{c_i^e}{c_i^\text{ref}}\right)^{\theta_i} \textit{exp}\left(\frac{\alpha_+^e n F \eta^e}{RT}\right) \\ -  \displaystyle \prod_{i \in \text{Ox.}} \left( \frac{c_i^e}{c_i^\text{ref}}\right)^{\theta_i} \textit{exp}\left(\frac{-\alpha_-^e n F \eta^e}{RT} \right)\Bigg].
\end{multline}

\noindent Here, \(j_0^e\) is the exchange current density and \(\alpha_\pm^e\) denotes an electrode's anodic (+) or cathodic (-) transfer coefficients. In a \mbox{steady-state} situation, the currents resulting from the anode's and cathode's current densities should be equal. Please note that the reference values for concentrations (i.e., \(c_i^\text{ref}\)) do not have to be the same for the Nernst equation and the Butler-Volmer equation.

\subsection{Fluid Dynamics}
Following from \cref{eq:Nernst,eq:ButlerVolmer}, the species concentrations play an important role in the fuel cell performance, and can be modeled through convective and diffusive transport in microfluidic devices. The fluid flow is modeled using the Hagen-Poiseuille law
\begin{equation}
\label{eq:HagenPoiseuille}
    \Delta p = Q \cdot R_h,
\end{equation}
\noindent where \(\Delta p\) is the pressure difference across a channel, \(Q\) is the volumetric flow rate, and \(R_h\) is the hydraulic resistance of a channel, for a given channel cross-section \cite{Oh2012}. 

With the obtained flow field, the concentration profile in a continuous phase can be modeled through the adapted advection-diffusion equation 
\begin{equation}
\label{eq:AdvectionDiffusionReduced}
    \frac{\partial^2 c_i}{\partial {y^*}^2} =\frac{u_x\,w}{D_i} \frac{\partial c_i}{\partial x^*},
\end{equation}
\noindent for electrically neutral species, where it is assumed that advection dominates the transport along the channel, and diffusion occurs only across the channel \cite{Wu2004,takken2025abstract}. Here, \(u_x\) is the average flow velocity, and \(D_i\) denotes the diffusion coefficient of species \(i\). For a flow channel with width \(w\), \(x^*\) and \(y^*\) denote the dimensionless coordinates in longitudinal and traversal direction of the flow channel, respectively, where \(x = x^*w\) and \(y = y^*w\). In case a species is electrically charged, the concentration profile follows from the Nernst-Planck equation, given by
\begin{equation}
\label{eq:NernstPlanckReduced}
    \frac{\partial^2 c_i}{\partial {y^*}^2} + \frac{z_i F}{R T} \left(\phi^C - \phi^A\right)\frac{\partial c_i}{\partial y^*} = \frac{u_x\,w}{D_i} \frac{\partial c_i}{\partial x^*}.
\end{equation}
\noindent Here, \(z_i\) is the valence of the ionic species and \(\phi^e\) is the electrode potential. Besides the assumptions for \cref{eq:AdvectionDiffusionReduced}, we additionally assumed here that external magnetic fluxes are nonexistent. 

Finally, the reaction rate for a species at an electrode is given by Faraday's law, i.e.,
\begin{equation}
\label{eq:FaradayLaw}
    R_i^e = \frac{A^e\,\theta_i \, j(c_i^e, \eta^e)}{n F}.
\end{equation}
\noindent Here, \(A^e\) is the active surface area, which indicates the ratio of the electrode surface with catalytically active sites. The reaction rate couples the boundary conditions of the concentration profiles back to the operating current density.

\begin{figure}[t]
    \centering
    \includegraphics[width=0.9\linewidth]{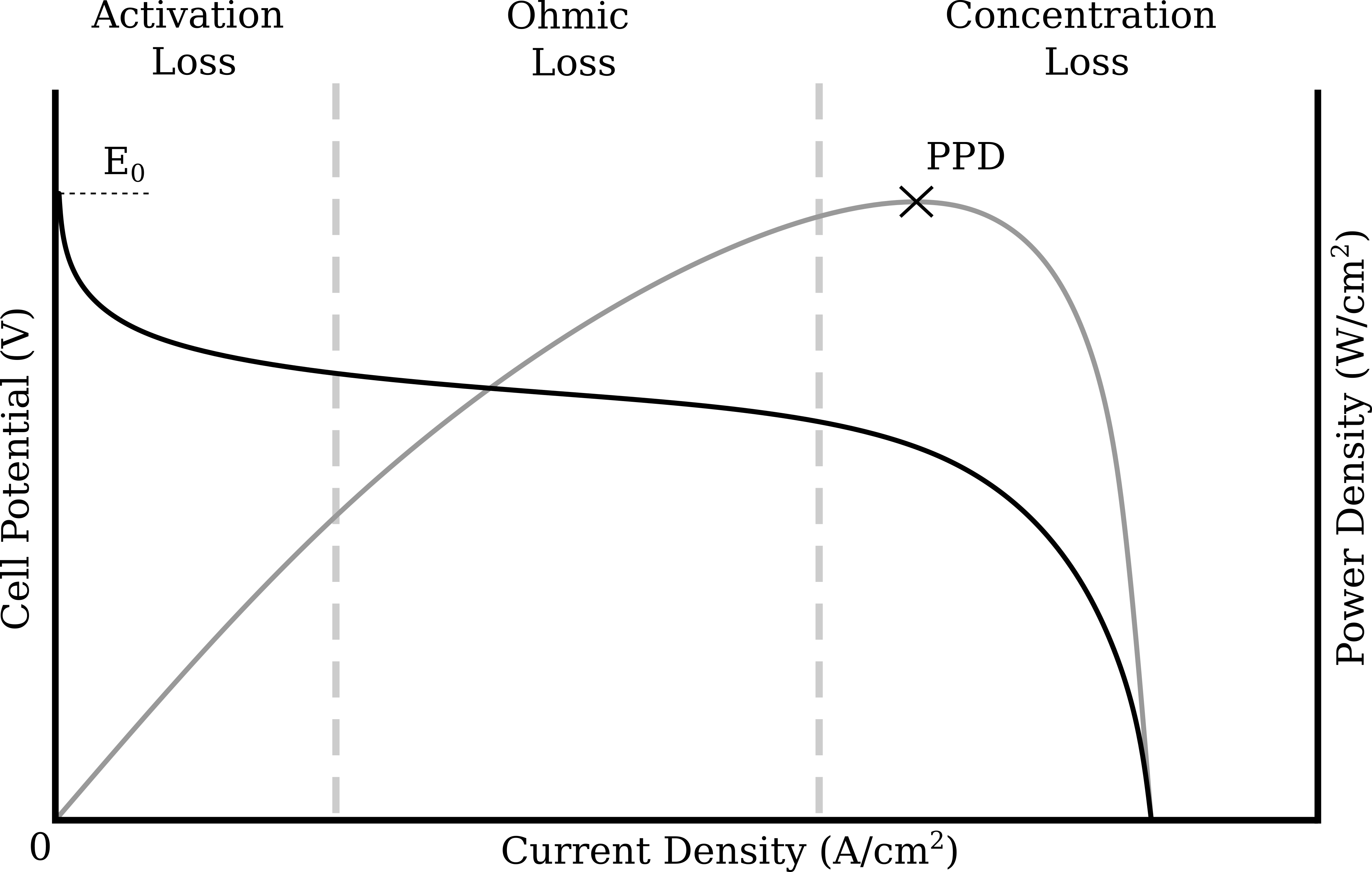}
    \caption{The polarization curve (black) is generally used to evaluate the performance of a fuel cell. The curve consists of three main losses: The overpotential loss, ohmic loss, and the mass transport loss. The power density curve (grey) follows from the polarization curve.}
    \label{fig:PolarizationCurve}
\end{figure}

\subsection{Polarization Curve}
Ultimately, the performance of a microfluidic fuel cell is evaluated using its \emph{polarization curve}, which relates the operating voltage to the current density. From this curve, the power density curve and key performance indicators such as the \emph{Peak Power Density}~(PPD) can be derived. Schematic examples of these curves are shown in \cref{fig:PolarizationCurve}. 

Besides providing performance indicators, the polarization curve also helps identify sources of performance loss, generally categorized into three types~\cite{Ohayre2016}. A fuel cell model should accurately capture all three losses:

\begin{itemize}
    \item The \emph{Activation Loss} is the loss that is caused by the required overpotential at the electrodes to drive electrochemical reactions. This loss is captured through \cref{eq:ButlerVolmer} and is typically reduced by using catalysts, such as platinum, at the electrodes surfaces.
    \item The \emph{Ohmic Loss} is caused by the ohmic resistance of the electrolyte and other cell components on the transport of charge carriers, such as ions. A reduction of this loss can be achieved by facilitating ionic transport across the electrolyte by, e.g., reducing the channel width. The ohmic loss in the electrolyte is modeled through \cref{eq:NernstPlanckReduced}.
    \item The \emph{Concentration Loss} is the loss caused by the scarcity of reactants, or abundance of reaction products, of the electrochemical reaction at high current density. Effective replenishment of reactants and removal of products is critical to minimize this loss. This loss is modeled through the transport of species and their reactions, captured by \cref{eq:AdvectionDiffusionReduced,eq:FaradayLaw}, respectively.
\end{itemize}

\noindent Each of these losses dominates the respective region of the polarization curve as illustrated in \cref{fig:PolarizationCurve}. Equipped with a resulting polarization curve and considering the respective losses, a design can be adapted to optimize the performance in terms of power by raising the PPD. However, these losses are coupled and design changes that reduce one loss often lead to an increase of another. Hence, accurate simulation for fuel cells is essential.

\section{Abstraction of a Microfluidic Fuel Cell}
\label{sec:SingleCell}
With the model from \cref{sec:Background} the behavior of a microfluidic fuel cell is understood and, hence, can be simulated. However, \mbox{state-of-the-art} methods using CFD are computationally expensive and scale badly \cite{Takken2022,Wu2024}. Hence, in this work, we propose an abstract method that exploits the simple geometry of microfluidic channels and, by this, allows for more efficient simulation, without losing accuracy. In this section, we describe how this is done for a single microfluidic fuel cell. Afterwards, the abstraction is extended to a microfluidic fuel cell stack in \cref{sec:Stack}.

\subsection{The Different Natures of Abstraction}
The mathematical model described in \cref{sec:Background} handles various physical phenomena, which can be abstracted differently. Here, we discuss the proposed abstraction method for the electrochemical effects, fluid dynamics, and species concentrations:

\begin{itemize}
    \item The equations related to electrochemistry, \cref{eq:cellPotential,eq:Nernst,eq:ButlerVolmer}, are explicit and can be solved easily once the variables are obtained. \cref{eq:ButlerVolmer} is solved using a root-finding method, to find the overpotential \(\eta^e\) for a given current density \(j\).
    \item The Hagen-Poiseuille equation, \cref{eq:HagenPoiseuille}, is an abstraction of the complex Navier-Stokes equations, defined on a simple geometry~\cite{Oh2012}.
    \item Conversely, the abstraction of the mathematical model of the concentration profile is rather related to the nature of its implementation, than an abstraction of the model itself. Whereas CFD methods solve governing equations on a discretized domain, we exploit the simplistic geometric nature of the problem and solve, in particular \cref{eq:AdvectionDiffusionReduced,eq:NernstPlanckReduced}, in a \mbox{pseudo-analytic} fashion. It involves the use of basis spaces to describe the concentration profiles, which require more in-depth treatment and are discussed next.
\end{itemize}

\begin{figure}[t]
    \centering
    \includegraphics[width=\linewidth]{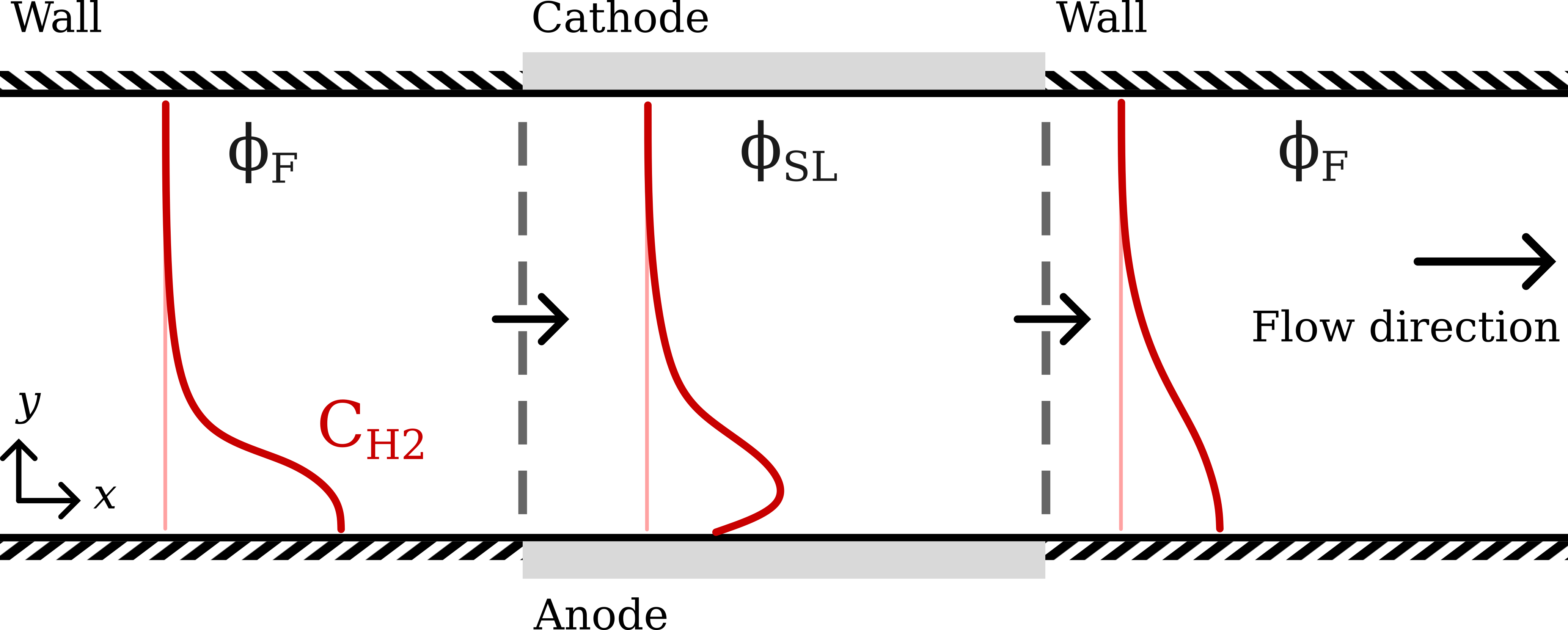}
    \caption{Schematic representation of a single microfluidic fuel cell with opposing anode and cathode of length \(L\). The concentration profile of hydrogen evolves as it flows through the channel. 
    }
    \label{fig:1D_representation}
\end{figure}

\subsection{Basis Function Spaces for Concentration Profiles}
\label{subsec:Cell}
A basis function space is a set of functions that can be linearly combined to represent a function. An example of such a basis space is the Fourier basis. To model the concentration distribution of species, following \cref{eq:AdvectionDiffusionReduced,eq:NernstPlanckReduced}, we make use of such basis spaces, and define inflowing concentration distributions in the corresponding basis. This basis can be constructed such that it solves for the evolution of the concentration distribution according to \cref{eq:AdvectionDiffusionReduced,eq:NernstPlanckReduced} and the corresponding local boundary conditions.

To this end, consider the flow channel depicted in \cref{fig:1D_representation} with a single cell consisting of an anode and an opposing cathode, separated by a flow channel. This cell is enclosed between two walled sections, i.e., sections where it is assumed that no electrochemical reactions occur. 
To simplify the discussion, we assume that the temperature is kept constant at a reference temperature by a thermal control subsystem,
and we focus on the concentration profile of \mbox{hydrogen \(C_{H_2}\)}, flowing in from the left. In the walled sections, the concentration profile of hydrogen evolves as it propagates through that section, due to diffusion. It does not undergo any changes due to electrochemical reactions at the boundary (walls). In fact, \cref{eq:AdvectionDiffusionReduced} in combination with these Neumann type boundary conditions, is the same as the heat equation with insulated ends problem \cite{BoyceDiPrima}. The solution to this problem is \mbox{well-known} and is found by performing separation of variables and solving the second order partial differential equation using the Fourier basis \(\phi_F\). As the profile propagates, the coefficients are manipulated based on the flow velocity, diffusion coefficient and eigenvalues of the problem, and the evolution of the concentration profile is given. As such, the concentration profile can be propagated through a microfluidic network \cite{takken2025abstract}.

In a similar fashion, the evolution of the concentration profile in presence of electrochemical reactions at the anode and/or cathode can be obtained. In this case, the electrochemical reaction incites a concentration flux at the electrode, and the resulting second order differential equations from \cref{eq:AdvectionDiffusionReduced,eq:NernstPlanckReduced} (after separation of variables) can be solved using the \mbox{Sturm-Liouville} basis \(\phi_{SL}\), following from the \mbox{Sturm-Liouville} theorem \cite{BoyceDiPrima}. 

\begin{figure*}[t]
    \centering
    \includegraphics[width=0.8\textwidth]{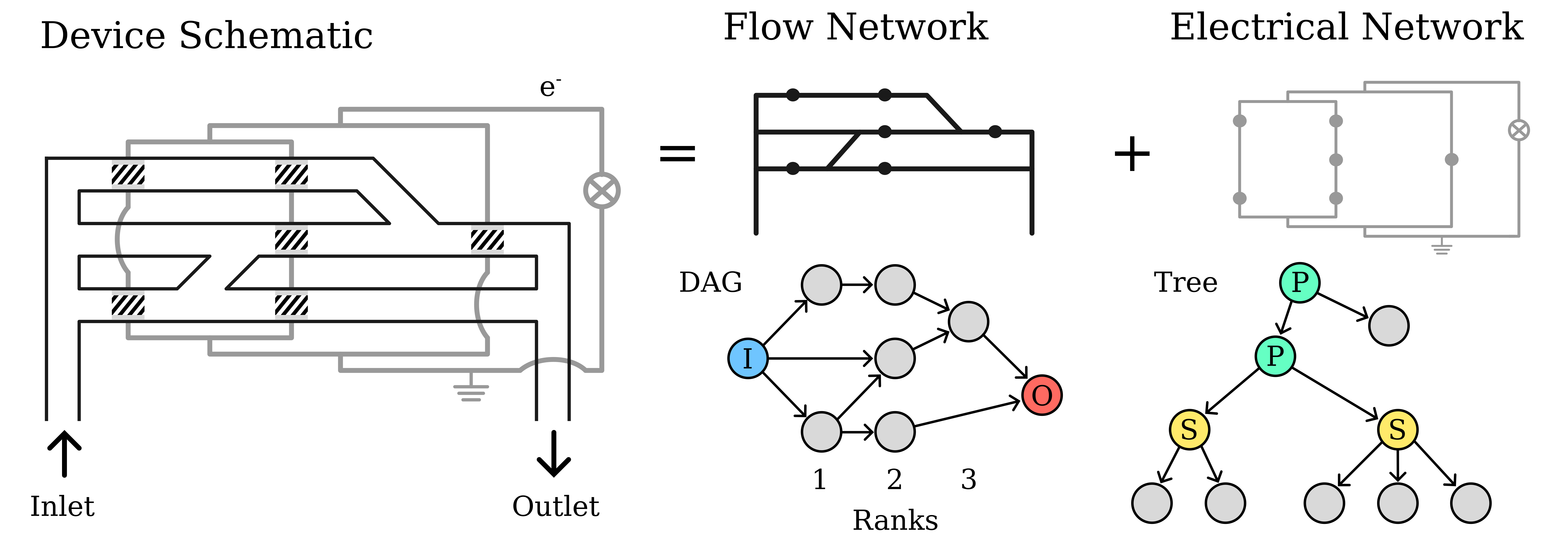}
    \caption{A microfluidic fuel cell stack can be separated into a flow network and an electrical network. The interface of these two networks is the set of fuel cells, which relate the operating current density with the fluid species concentrations, and vice versa. In each of these networks, the cells can be connected in serial, as well as parallel.}
    \label{fig:NetworkSplit}
\end{figure*}

\subsection{Boundary Condition Models}
\label{sec:BCModels}
The boundary conditions specify the behavior of the concentration profile at the boundaries (walls and electrodes) of the channel. They play a significant role in the definition of the basis function spaces and must be defined carefully. The formal definition of the boundary conditions for \cref{eq:AdvectionDiffusionReduced,eq:NernstPlanckReduced} is given by 
\begin{equation}
\label{eq:BoundaryConditions}
    \frac{\partial c_i}{\partial y}\bigg\vert_{W} = 0, \quad \text{and}\quad \frac{\partial c_i}{\partial y}\bigg\vert_{E} = \frac{R^e_i}{D_i},
\end{equation}

\noindent where \(W\) denotes the wall section and \(E\) the electrode section. These are both of Neumann type and, unfortunately, the \mbox{Sturm-Liouville} problem with Neumann type boundary conditions that are non-zero is not self-adjoint. To circumvent this problem, we model the boundary condition for the electrode section as
\begin{equation}
\label{eq:BoundaryConditions}
    \frac{\partial c_i}{\partial y}\bigg\vert_{E} = \frac{q^e_i}{D_i} \cdot c_i^e, \qquad \text{where }\,\,\, q_i^e = \frac{A^e\,\theta_i \,j(c_i^e, \eta^e)}{n F \tilde{c}_i^e}.
\end{equation}

\noindent Here, \(\tilde{c}_i\) is an approximation of the concentration value at the electrode boundary. Using this approximation, we obtain a \mbox{quasi-Robin} type boundary condition and the Sturm-Liouville problem becomes self-adjoint. The definition of \(\tilde{c}_i\) plays a significant role in the meaningfulness of the results. In this work, the value of \(\tilde{c}_i\) is evaluated iteratively, until the change in concentration profile is consistent with the reaction rate \(R_i^e\). 

Overall, the proposals listed above provide an abstract (yet accurate) description of the behavior of microfluidic fuel cells that can be simulated much more efficiently. The accuracy of this boundary condition model is evaluated in \cref{sec:Justification} for a single cell. However, to provide a full overview of the proposed method, the extension of the abstraction to microfluidic fuel cell stacks is first discussed next.

\vspace{7mm}

\section{Extension to a Microfluidic Fuel Cell Stack}
\label{sec:Stack}
So far, the discussion has focused on a single cell microfluidic fuel cell. In this section, we extend the simulation approach to a fuel cell stack. To this end, consider the schematic representation of a stack device as shown in \cref{fig:NetworkSplit}. The device consists of a flow network and an electrical network that intersect at six nodes corresponding to six individual fuel cells. The two networks are first treated separately, after which the overall solution scheme for the complete system is presented.

\vspace{3mm}

\subsection{Flow Network}
Each individual cell operates at its local polarization curve. However, since the operating current density strongly depends on the species concentrations (see \cref{eq:ButlerVolmer}), each cell strongly depends on the state of the cells that lie upstream in the flow network, as those consume or produce species. Representing the flow network as a \emph{Directed Acyclic Graph} (DAG), as shown in \cref{fig:NetworkSplit}, provides a flow hierarchy, where the root node is the inflow and the outflow is the final node. Based on this, the fuel cells can be ranked according to their level of dependency on fuel cells upstream, providing an order in which the species should be propagated and solved. With this, and the process described in \cref{subsec:Cell}, the operating current density of each individual cell can be obtained locally.

\vspace{3mm}

\subsection{Electrical Network}
The fuel cells in \cref{fig:NetworkSplit} are connected with a set of series and parallel connections in the electrical network. This is represented in \cref{fig:NetworkSplit} with a \emph{Tree} graph, where the green nodes have children that are connected in parallel, and yellow nodes have children that are connected in series. The platinum leaf nodes are here fuel cells, but can also be other electrical components such as resistors. We wish to solve the electrical network and obtain the unknown currents and electrode overpotentials in the system, which will be stored in a vector of unknowns. The system of equations to model the electrical network is established using the Kirchhoff laws and Ohm's law, in addition to \cref{eq:ButlerVolmer}, which gives the operating current and overpotential for a fuel cell electrode. The system is \mbox{non-linear} and can be solved iteratively using the Newton-Raphson method, i.e.,

\begin{align}
\label{eq:NewtonRaphson1}
\bm{J}\left( \bm{u}^{(k)} \right)\cdot \Delta\bm{u}^{(k)} &= -\bm{f}\left(\bm{u}^{(k)}\right) \quad \text{and}\\
\label{eq:NewtonRaphson2}
\bm{u}^{(k+1)} &= \bm{u}^{(k)} + \Delta \bm{u}^{(k)},
\end{align}

\noindent where \(\bm{u}, \, \bm{f} \; \in \mathbb{R}^N \) denote the unknown vector and the functions vector, respectively, and \( \bm{J} \; \in \mathbb{R}^{N\times N } \) is the Jacobian of the system. The system size \(N\) follows from the sum of unknown currents and unknown electrode overpotentials in the system.

\newpage
\subsection{Overall System and Algorithm}
Given the DAG and Tree described above, the stack potential for a pre-defined stack operating current can be obtained iteratively, following the algorithm outlined in \cref{alg:SolvingScheme}. From this the stack polarization curve, and hence the peak power of a microfluidic fuel cell stack, is obtained. This eventually results in a simulation method that was implemented and tested for various large scale microfluidic fuel cell stacks. The performance of this simulation method is further discussed in \cref{sec:Performance}. However, first we discuss how it was validated that the method provides accurate results.

\begin{algorithm}[H]
\caption{Stack Solving Scheme}
\begin{algorithmic}[1]
\label{alg:SolvingScheme}
\renewcommand{\algorithmicrequire}{\textbf{Input:}}
\renewcommand{\algorithmicensure}{\textbf{Output:}}
\REQUIRE{\textit{I\_max; DAG G(V, E); Tree T(N, B)}}
\ENSURE{\textit{Stack potentials}}
\STATE Solve flow field
\FOR {I = 0 ... I\_max}
\WHILE{not converged}
\FOR{rank \(\in\) G}
\STATE Propagate species
\ENDFOR
\WHILE{res{ }\textless{ }\textepsilon}
    \STATE Solve Newton-Raphson
\ENDWHILE
\STATE Check convergence.
\ENDWHILE
\STATE Store stack potential.
\ENDFOR
\end{algorithmic}
\end{algorithm}

\begin{figure}[t]
    \centering
    \includegraphics[width=0.9\linewidth]{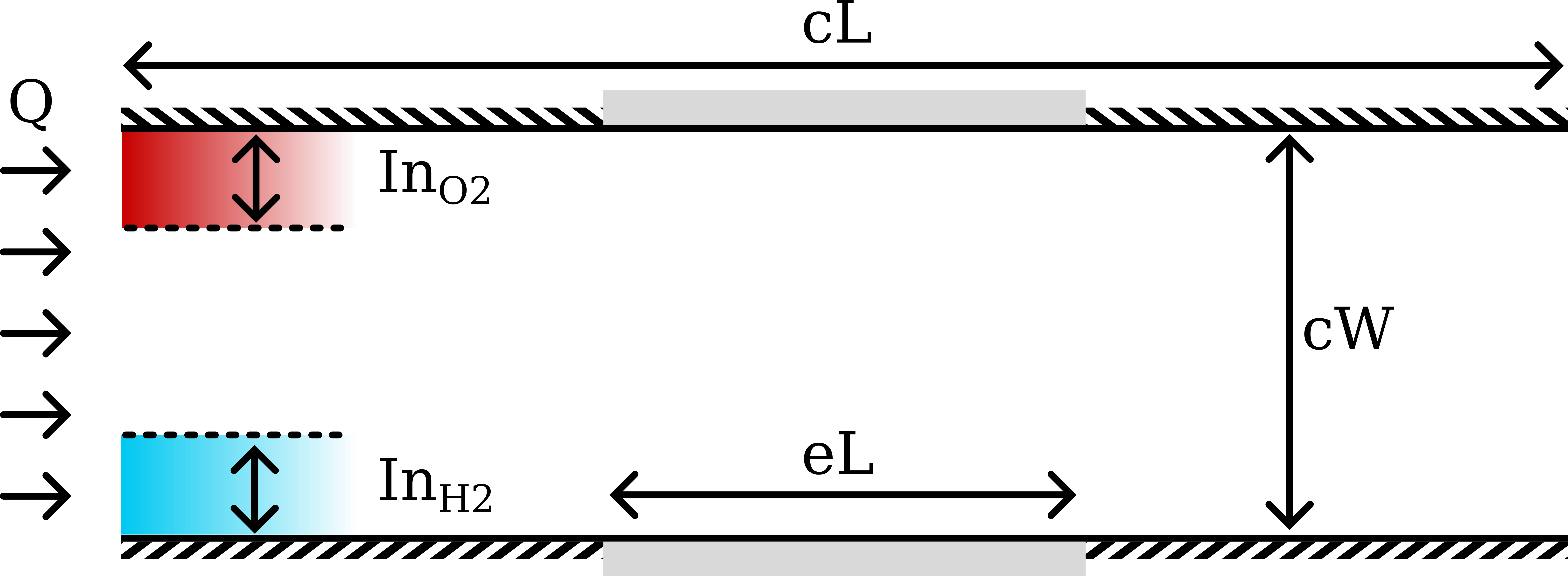}
    \caption{The single cell channel used as validation case.}
    \label{fig:validationCase}
\end{figure}

\begin{table}[t]
    \centering
    \caption{Geometric Parameters}
    \label{tab:geometryparam}
    \def\arraystretch{1.2}
    \begin{tabular}{@{\,}l@{\qquad}c@{\quad}r}
			  \textbf{Variable}& \textbf{Unit} & \multicolumn{1}{c}{\textbf{Value}} \\
			\midrule
            cL & \(m\) & \(1.0\cdot 10^{-3}\) \\
            cW & \(m\) & \(1.0\cdot 10^{-4}\) \\
            cH* & \(m\) & \(1.0\cdot 10^{-4}\) \\
            eL & \(m\) & \(1.0\cdot 10^{-3}\) \\
			\bottomrule
            \multicolumn{3}{l}{* \footnotesize Used to reduce dimensionality.}
	\end{tabular}
\end{table}

\begin{table}[t]
    \centering
    \caption{Flow parameters}
    \label{tab:globalparam}
    \def\arraystretch{1.2}
    \begin{tabular}{@{\,}l@{\qquad}c@{\quad}l@{\quad}l}
			  \textbf{Variable}& \textbf{Unit} & \multicolumn{1}{c}{\textbf{Value}} & \\
			\midrule
            \(A^e\) & - & \(10^{3}\) \\
            In\textsubscript{H2} & - & \(0.1\) \\
            In\textsubscript{O2} & - & \(0.1\) \\
            \(Q\) & \(\frac{m^3}{s}\) & \(1.0\cdot 10^{-9}\) & \\
            \(C_{H_2}\) & \(\frac{mol}{L}\) &  \(0.1\) & \\
            \(C_{O_2}\) & \(\frac{mol}{L}\) &  \(0.1\) & \\
            \(C_{OH}\) & \(\frac{mol}{L}\) &  \(1.0\) & \\
            \(C_{H_2O}\) & \(\frac{mol}{L}\) &  \(55.4\) & \\
            \(D_{H_2}\) & \(\frac{cm^2}{s}\) &  \(5.1324\cdot 10^{-5}\) & \cite{tham1970diffusion} 
            \\
            \(D_{O_2}\) & \(\frac{cm^2}{s}\) &  \(2.0094\cdot 10^{-5}\) & \cite{tham1970diffusion} 
            \\
            \(D_{OH}\) & \(\frac{cm^2}{s}\) &  \(2.6880\cdot 10^{-5}\) & \cite{see1998diaphragm}\\
            \(D_{H_2O}\) & \(\frac{cm^2}{s}\) &  \(2.2990\cdot 10^{-5}\) & \cite{mccall1965effect} 
            \\
			\bottomrule
	\end{tabular}
\end{table}

\begin{table}[t]
    \centering
    \caption{Electrochemical Parameters}
    \label{tab:electrochemparam}
    \def\arraystretch{1.7}
    \begin{tabular}{@{\,}l@{\qquad}c@{\quad}l@{\quad}c@{\quad}l@{\quad}c}
            & & \multicolumn{4}{c}{\textbf{Value}}\\
            \cmidrule{3-6}
			  \textbf{Variable} & \textbf{Unit} & \multicolumn{2}{c}{\textbf{Nernst Eq.}} & \multicolumn{2}{c}{\textbf{Butler-Volmer Eq.}} \\
			\midrule
			\(E^0_{A,\text{ref}}\) & \(V\) & -0.8277 & \cite{vanysek2000electrochemical} & - & \\
            \(E^0_{C,\text{ref}}\) & \(V\) & 0.401 & \cite{vanysek2000electrochemical} & - & \\
            \(j_0^A\) & \(\frac{A}{cm^2}\) & - & & \(6.743\cdot10^{-4}\) & \cite{Sheng2010hydrogenAnode}\\
            \(j_0^C\) & \(\frac{A}{cm^2}\) & - & & \(4.2\cdot10^{-11}\) & \cite{Blurton1972lowCathode}\\
            \(\alpha_+^A\) & - & - & & 0.638995 & \cite{Kunz2019modeling} \\
            \(\alpha_+^C\) & - & - & & 0.595925 & \cite{Kunz2019modeling} \\
            \(\alpha_-^A\) & - & - & & 0.361005 & \cite{Kunz2019modeling} \\
            \(\alpha_-^C\) & - & - & & 0.404075 & \cite{Kunz2019modeling} \\
            \(c_{H_2}^{0,A}\) & \(\frac{mol}{L}\) & 0.59705 & \cite{Kunz2019modeling} & 0.77612 & \cite{Kunz2019modeling}\\
            \(c_{OH}^{0,A}\) & \(\frac{mol}{L}\) & 1.0 & \cite{vanysek2000electrochemical} & 100.0 & \cite{Sheng2010hydrogenAnode}\\
            \(c_{H_2O}^{0,A}\) & \(\frac{mol}{L}\) & 54.918 & \cite{Kunz2019modeling} & 55.373 & \cite{Kunz2019modeling}\\
            \(c_{O_2}^{0,C}\) & \(\frac{mol}{L}\) & 0.85206 & \cite{Kunz2019modeling} & 0.083457 & \cite{Kunz2019modeling}\\
            \(c_{OH}^{0,C}\) & \(\frac{mol}{L}\) & 1.0 & \cite{vanysek2000electrochemical} & 6.880 & \cite{Blurton1972lowCathode}\\
            \(c_{H_2O}^{0,C}\) & \(\frac{mol}{L}\) & 54.918 & \cite{Kunz2019modeling} & 49.982 & \cite{Kunz2019modeling}\\
			\bottomrule
	\end{tabular}
\end{table}

\section{Validation of the Abstraction}
\label{sec:Justification}
To confirm that the made abstractions are valid and, hence, the proposed method works, we validated them using results obtained from a CFD simulation. To this end, we considered a fuel cell stack with a single cell 
and the flow network from \cref{fig:NetworkSplit}. This section first describes these validation cases. Afterwards, we discuss the obtained results.

\subsection{Validation Case: Single Cell}
The single cell system that is considered for the validation is illustrated in \cref{fig:validationCase}. Similar to the cell in \cref{fig:1D_representation}, this cell is a flow channel containing an anode, cathode, and isolating walls. The geometric parameters are given in \cref{tab:geometryparam}, where cL, cW, cH, and eL are the channel length, channel width, channel height, and electrode length, respectively. Here, it should be noted that the simulation is performed in a 2-D domain and the depth parameter, cH, is used to reduce the dimensionality of other parameters (e.g., concentration values, which are given in volumetric units).

At the channel inlet on the left, the concentrations of water and hydroxide are uniform, whereas hydrogen and oxygen have a concentration distribution in the form of a step-function. This is illustrated in \cref{fig:validationCase} by the blue and red domains, which indicate what channel region contains concentrated hydrogen or oxygen, respectively. The ratio of the channel width that is filled with the corresponding species, is given by In\textsubscript{H2} and In\textsubscript{O2}. Otherwise, there is no hydrogen or oxygen entering the channel. The species flow in with a flow rate \(Q\), and the concentration at inflow and diffusion of the species are given by \(C_i\) and \(D_i\), respectively, which are all listed in \cref{tab:globalparam}.

The electrochemical parameters that are needed to solve \cref{eq:Nernst,eq:ButlerVolmer} are provided in \cref{tab:electrochemparam}. These parameters, together with the diffusion coefficients \(D_i\), were taken from literature on experiments (sources are listed in \cref{tab:globalparam,tab:electrochemparam}).

\subsection{Validation Case: Flow Network}
\begin{figure}[t]
    \centering
    \includegraphics[width=1.0\linewidth]{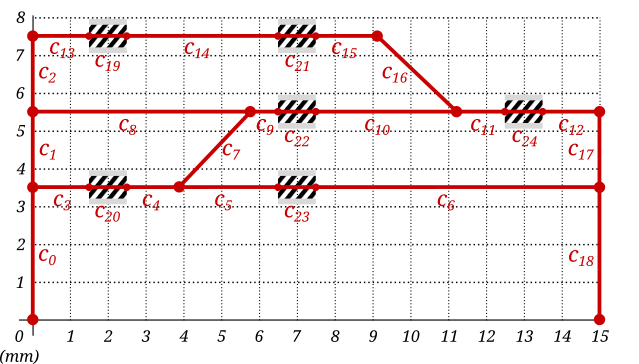}
    \caption{A detailed schematic representation of the flow network illustrated in \cref{fig:NetworkSplit}. All channels have a width and height of 0.1 mm, and are labeled \(c_0\) through \(c_{24}\), where the six fuel cells are denoted by \(c_{19}\) through \(c_{24}\).}
    \label{fig:FlowNetwork}
\end{figure}

To evaluate the accumulated error introduced by the propagation model illustrated in \cref{fig:1D_representation}, we validate the proposed method using the complex flow network from \cref{fig:NetworkSplit}, which is illustrated in more detail in \cref{fig:FlowNetwork}. Although this network is primarily intended for illustrative purposes and is not representative of a practical configuration, it includes important flow operations such as, idle flow, splitting and merging of streams, combined with flow through operating fuel cells. Evaluation of the propagation through this network will provide useful information on the proposed method's quality.

The inlet stream ratios In\textsubscript{H2} and In\textsubscript{O2} are set to 1.0, to maintain stability while allowing a reduced \mbox{mesh-resolution} in the CFD solver. Although this does not represent a typical setup, as it
introduces crossover currents, the computational analysis of the proposed method is
unaffected by this choice. The inflow and outflow are located on the left- and right-hand side in \cref{fig:FlowNetwork}, respectively, and the validation is conducted at three different inflow velocities: 1, 10, and 100 mm/s, whose results are analyzed in the following section.

\begin{figure}[t]
    \centering
    \includegraphics[width=1.0\linewidth]{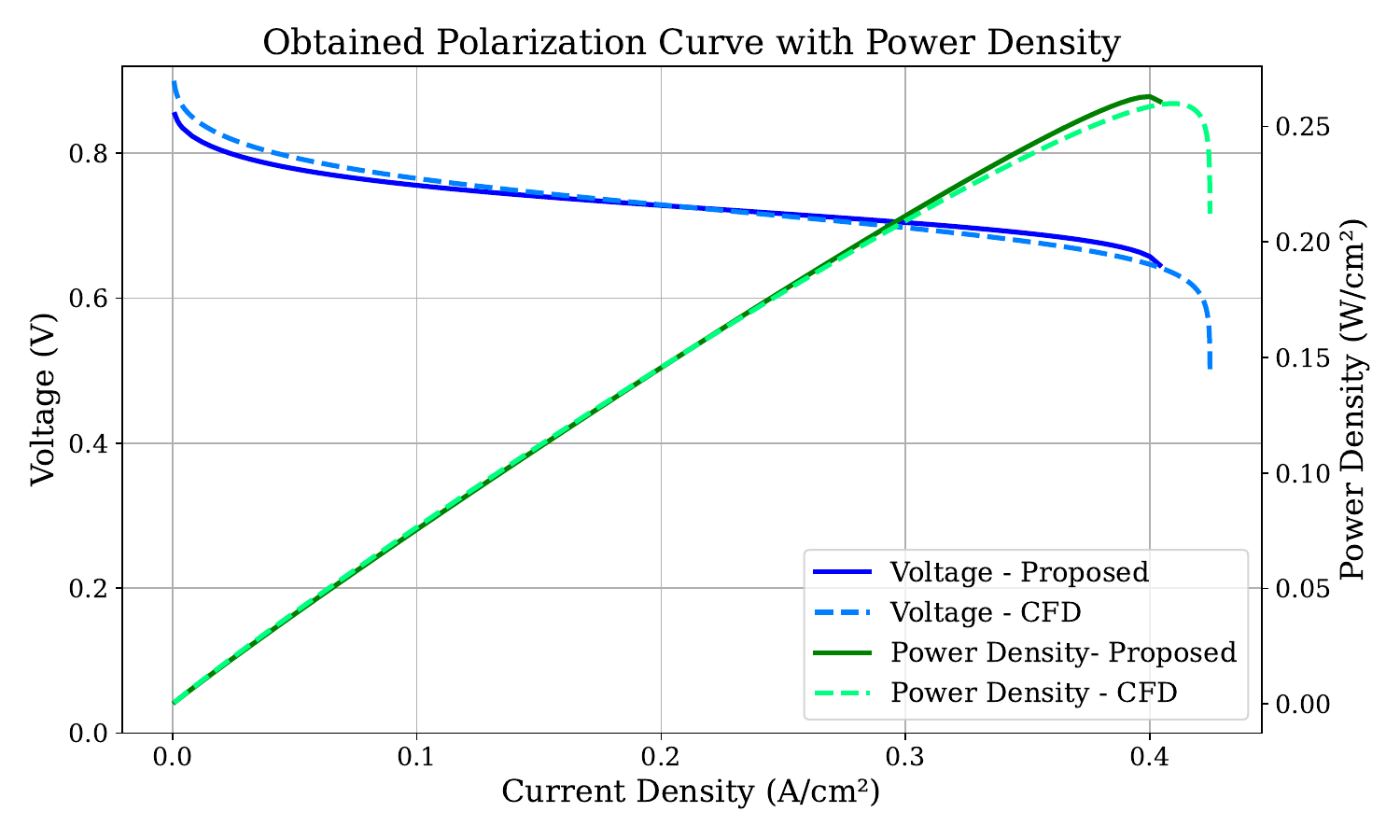}
    \caption{The polarization and power density curves for the validation case as obtained from the CFD simulation and the proposed method.}
    \label{fig:PolarizationCurveResult}
\end{figure}

\begin{figure}[t]
    \centering
    \begin{minipage}{1.0\columnwidth}
        \centering
        \begin{subfigure}[b]{1.0\columnwidth}
        \centering
            \includegraphics[width=\linewidth]{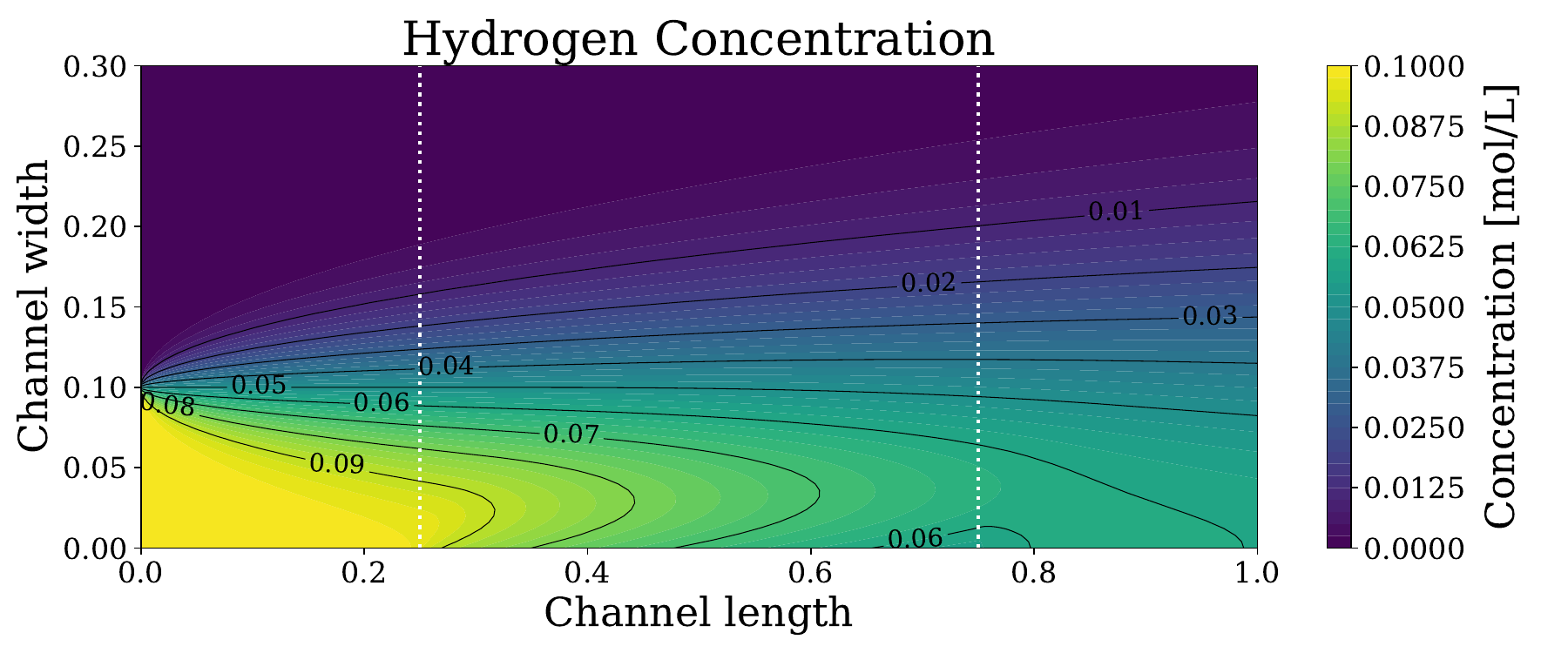}
            \caption{Proposed}
            \label{fig:hydrogenAbstract}
        \end{subfigure}
        \begin{subfigure}[b]{1.0\columnwidth}
            \centering
            \vspace{3mm}
            \includegraphics[width=\linewidth]{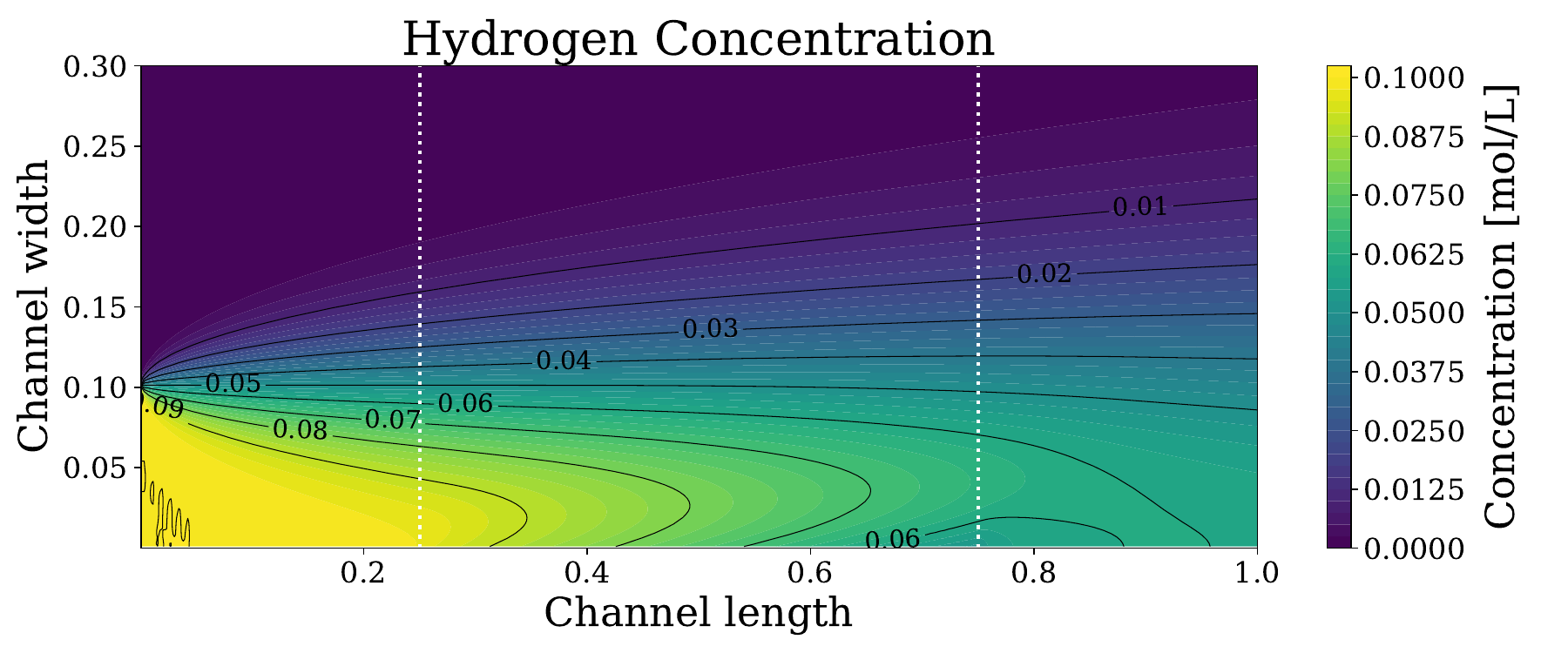}
            \caption{CFD}
            \label{fig:hydrogenCFD}
        \end{subfigure}
        \caption{The hydrogen distribution in the bottom 30\% of the channel.}
        \label{fig:concentration_profiles_hydrogen}
    \end{minipage}\hfill\\
    \vspace{5mm}
    \begin{minipage}{1.0\columnwidth}
        \centering
        \begin{subfigure}[b]{1.0\columnwidth}
        \centering
            \includegraphics[width=\linewidth]{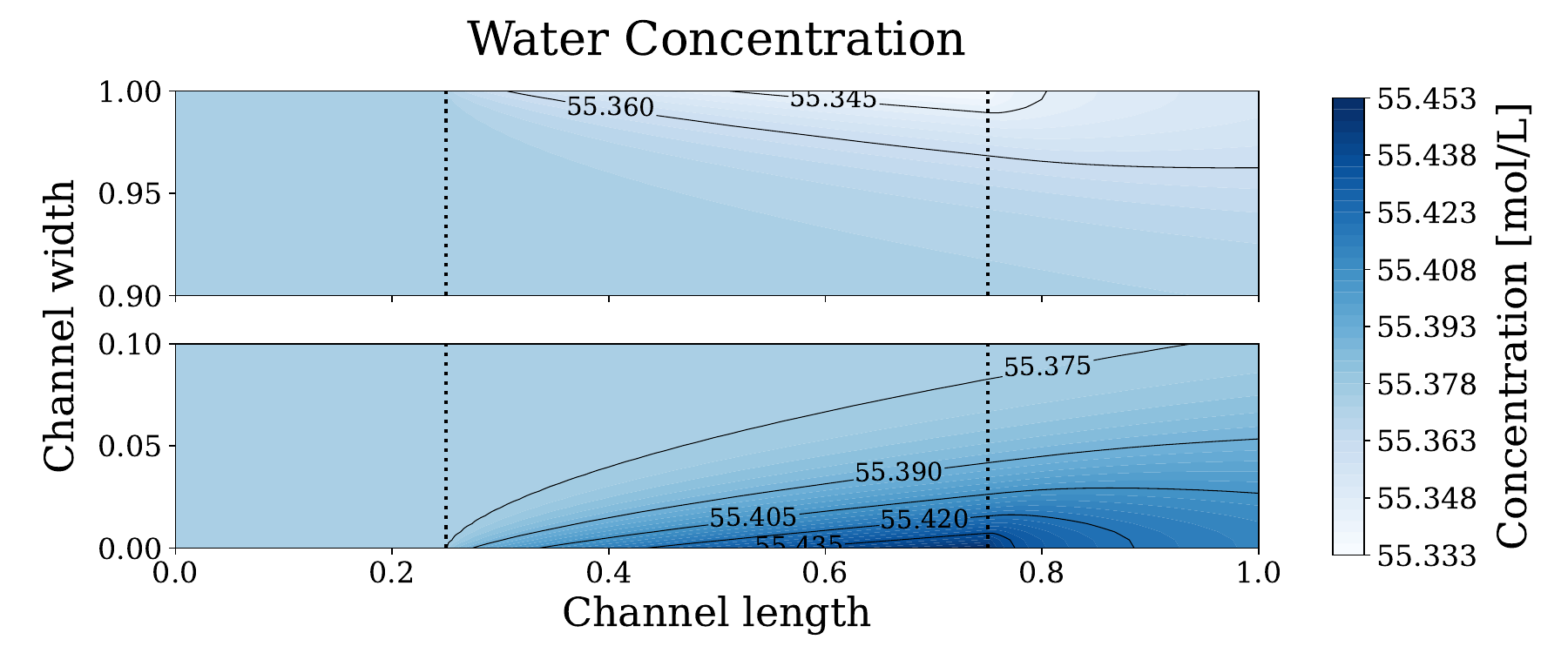}
            \caption{Proposed}
            \label{fig:waterAbstract}
        \end{subfigure}
        \begin{subfigure}[b]{1.0\columnwidth}
            \centering
            \vspace{3mm}
            \includegraphics[width=\linewidth]{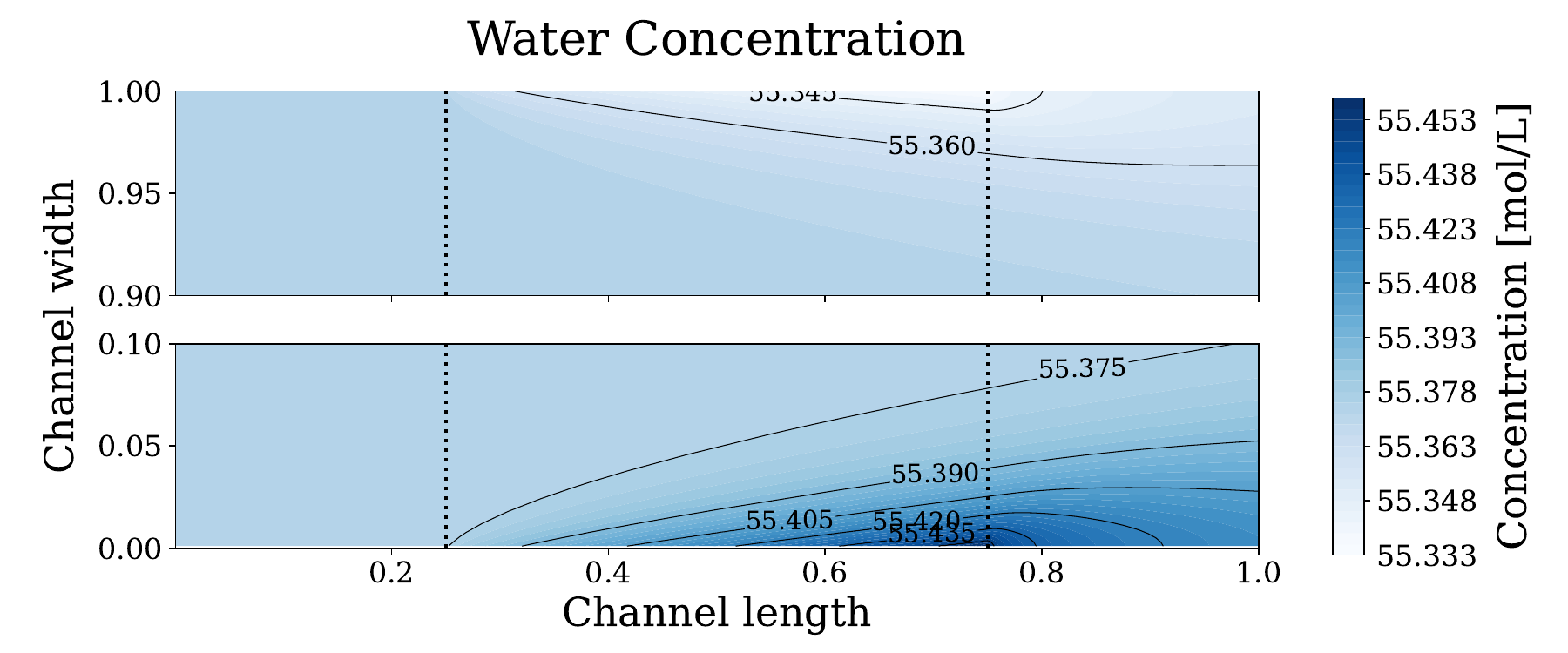}
            \caption{CFD}
            \label{fig:waterCFD}
        \end{subfigure}
        \caption{The water distribution in the top 10\% and bottom 10\% of the channel.}
        \label{fig:concentration_profiles_water}
    \end{minipage}
\end{figure}

\begin{figure}[t]
    \centering
    \begin{minipage}{1.0\columnwidth}
        \centering
        \begin{subfigure}[b]{1.0\columnwidth}
        \centering
            \includegraphics[width=\linewidth]{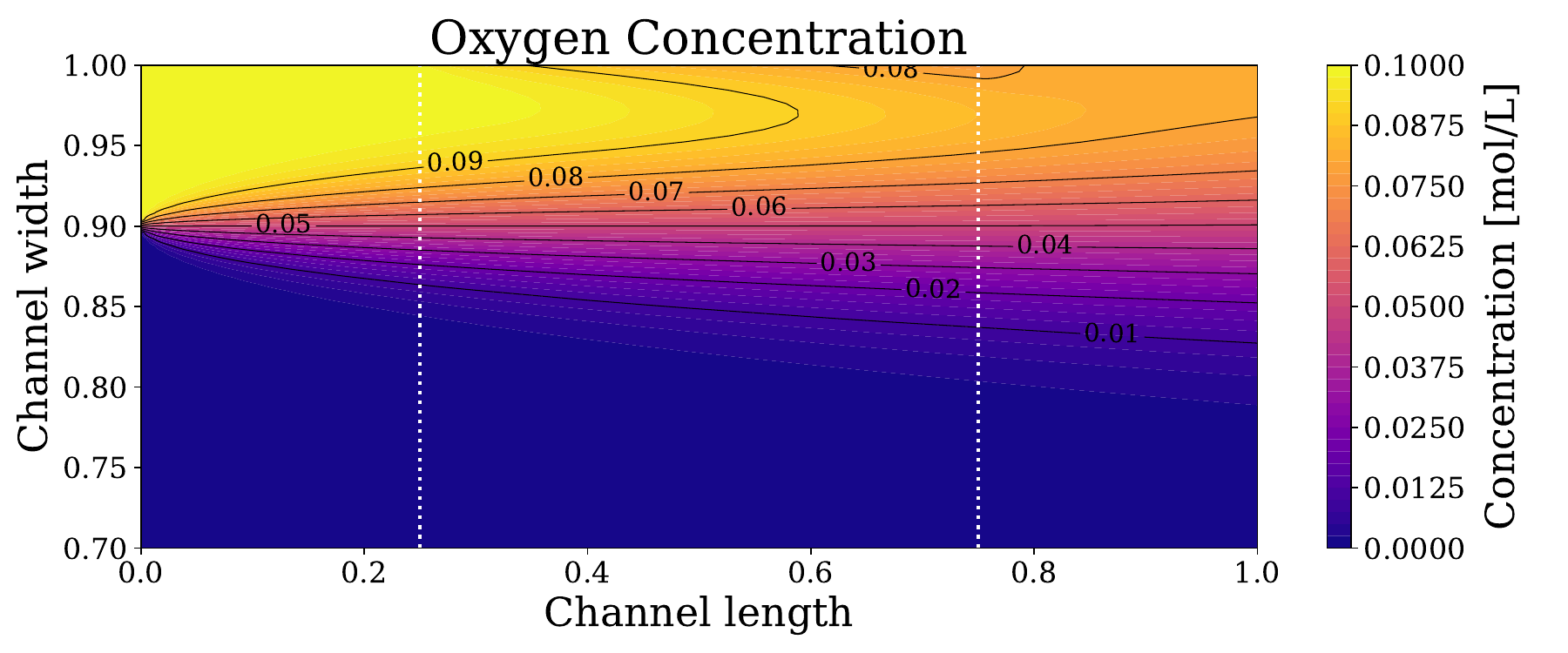}
            \caption{Proposed}
            \label{fig:oxygenAbstract}
        \end{subfigure}
        \begin{subfigure}[b]{1.0\columnwidth}
            \centering
            \vspace{3mm}
            \includegraphics[width=\linewidth]{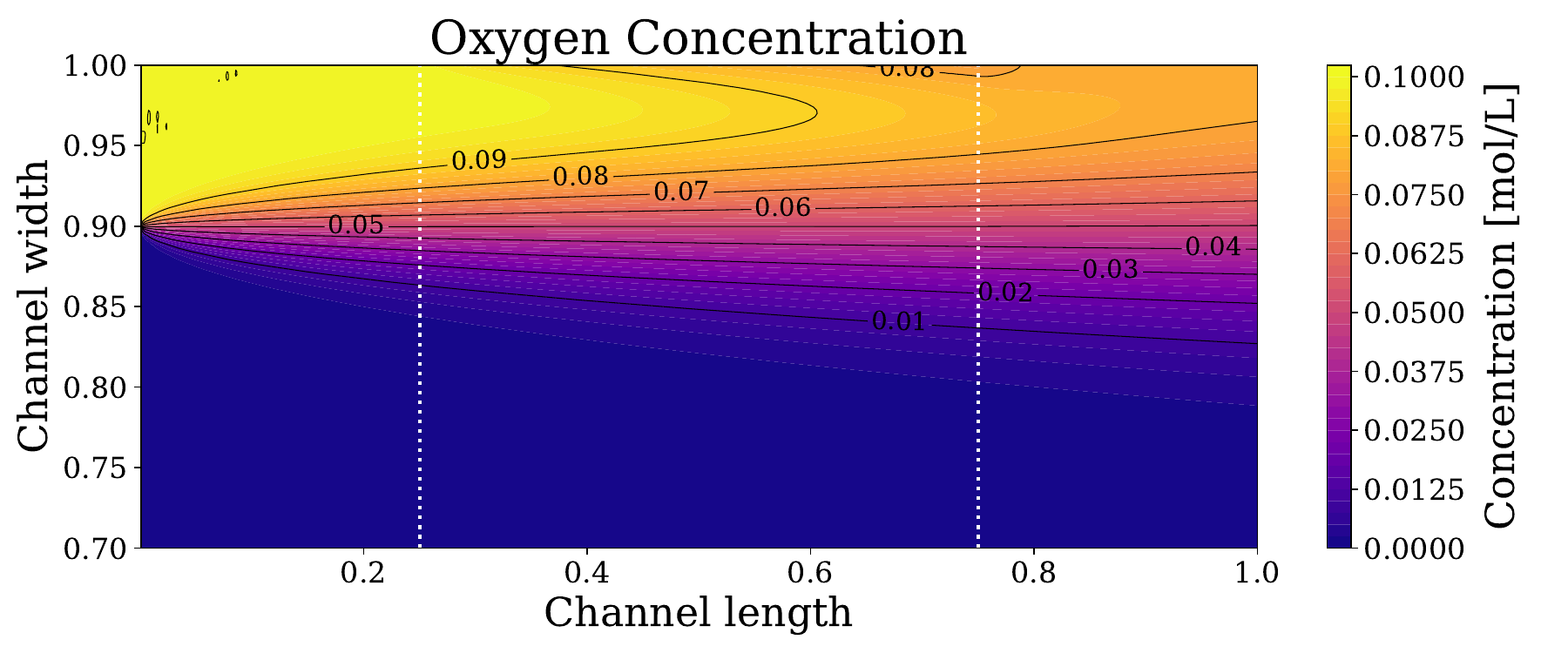}
            \caption{CFD}
            \label{fig:oxygenCFD}
        \end{subfigure}
        \caption{The oxygen distribution in the top 30\% of the channel.}
        \label{fig:concentration_profiles_oxygen}
    \end{minipage}\hfill\\
    \vspace{5mm}
    \begin{minipage}{1.0\columnwidth}
        \centering
        \begin{subfigure}[b]{1.0\columnwidth}
        \centering
            \includegraphics[width=\linewidth]{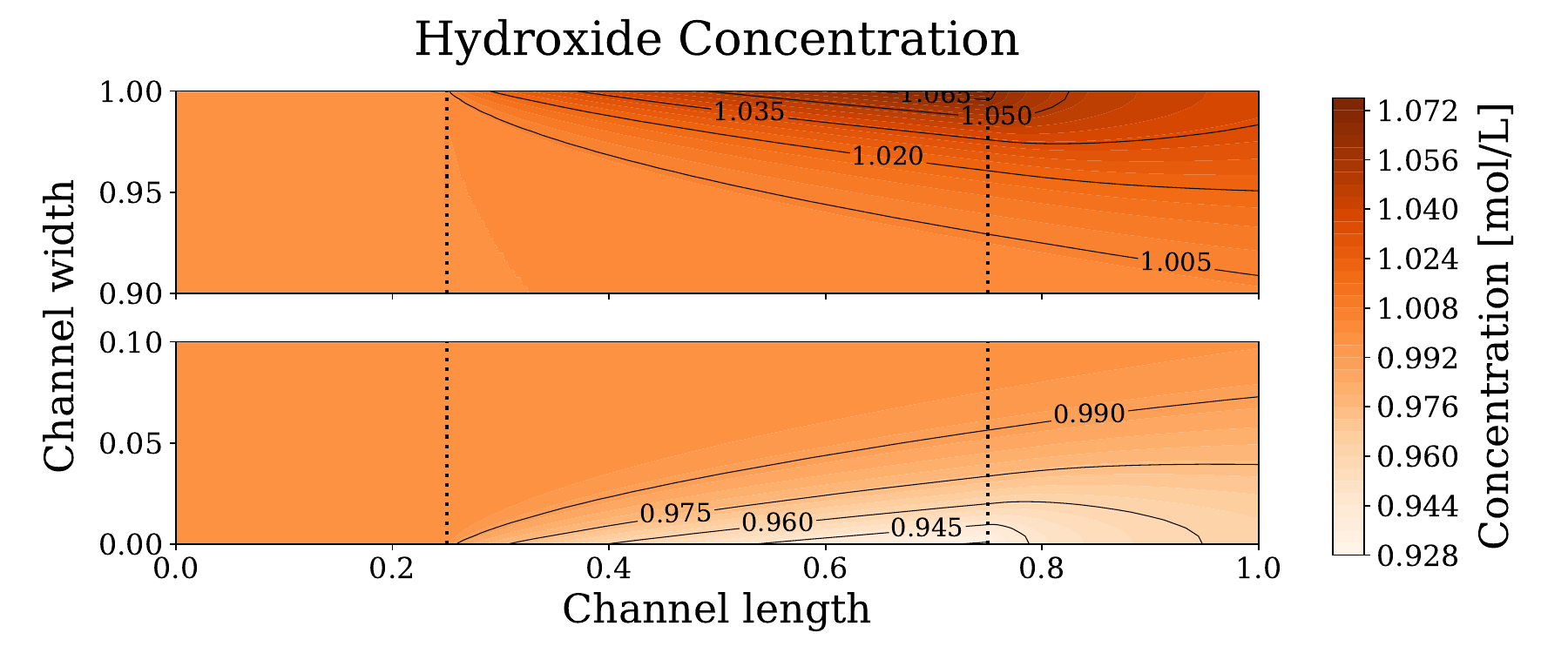}
            \caption{Proposed}
            \label{fig:waterAbstract}
        \end{subfigure}
        \begin{subfigure}[b]{1.0\columnwidth}
            \centering
            \vspace{3mm}
            \includegraphics[width=\linewidth]{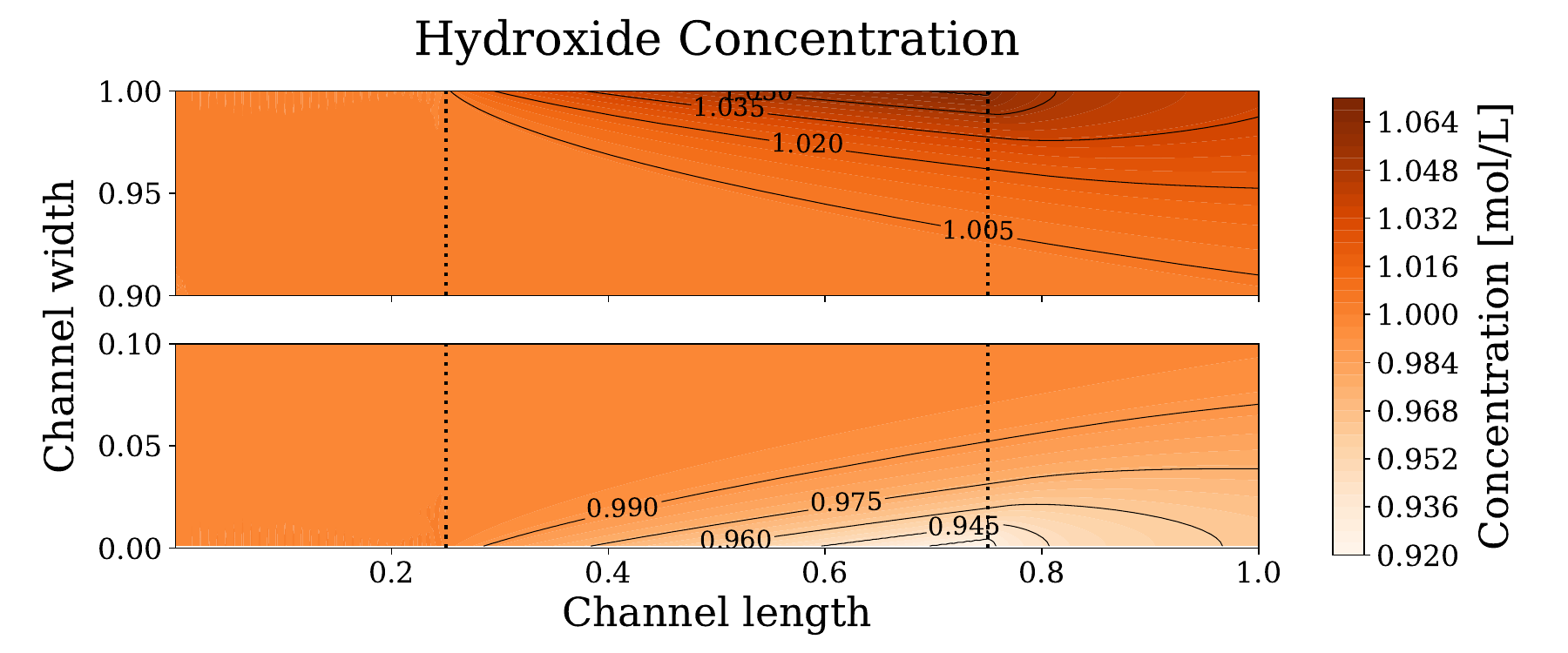}
            \caption{CFD}
            \label{fig:waterCFD}
        \end{subfigure}
        \caption{The hydroxide distribution in the top 10\% and bottom 10\% of the channel.}
        \label{fig:concentration_profiles_hydroxide}
    \end{minipage}
\end{figure}

\begin{figure*}[t]
    \centering
    \includegraphics[width=0.9\linewidth]{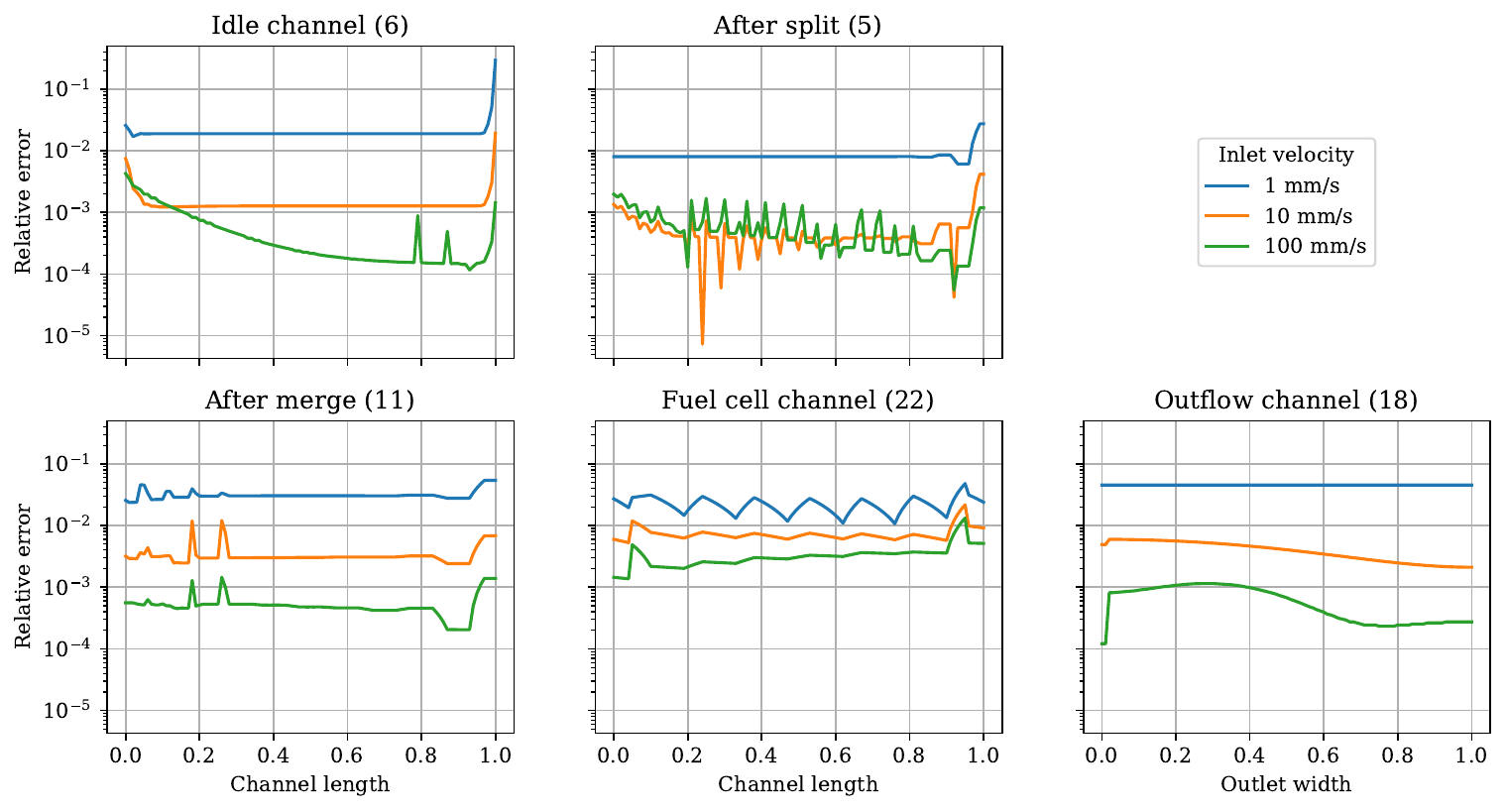}
    \caption{The relative errors for relevant propagation operations and at the outlet, for inflow velocities: 1, 10, and 100 mm/s. The error in average concentration was evaluated along representative channels (with the corresponding label provided in between brackets) for the four operations: idle flow, stream split, stream merge, and an operating fuel cell. The relative error across the outlet width is visualized.}
    \label{fig:PropagationError}
\end{figure*}

\subsection{Obtained Results}
As baseline for the validation, we used CFD simulation results which were obtained using COMSOL Multiphysics\circledR~\cite{comsol}. The resulting polarization curve for the single-cell validation case is given in \cref{fig:PolarizationCurveResult} for CFD as well as for the proposed method. Those results clearly show that the curve resulting from the proposed method is well in line with the curve obtained from CFD results. 

To confirm that the basis function spaces, that were introduced in \cref{sec:SingleCell}, accurately simulated the concentration distributions, the species concentration distributions obtained using both methods are shown in \cref{fig:concentration_profiles_hydrogen,fig:concentration_profiles_oxygen,fig:concentration_profiles_water,fig:concentration_profiles_hydroxide}, at an operating current density of 0.1~\mbox{A/cm\textsuperscript{2}}. The spatial dimension is given as the ratio to the channel width or length. The white or black dotted lines, positioned at 25\% and 75\% of the channel length, indicate the electrode-wall interfaces and the electrode is positioned in the middle, similar to \cref{fig:1D_representation}. The concentration distributions in \cref{fig:concentration_profiles_hydrogen,fig:concentration_profiles_oxygen,fig:concentration_profiles_water,fig:concentration_profiles_hydroxide} are shown with a focus on the relevant section of the channel width. The concentration profiles obtained using basis function spaces from the proposed method are well in line with those obtained from the CFD simulation. 

The evaluation of concentration propagation through the four relevant operations was done using representative channels from the flow network. Following the labeling from \cref{fig:FlowNetwork}:
\begin{itemize}
    \item channel 6 was used as representative of idle flow,
    \item channel 5 for flow after stream split,
    \item channel 11 for flow after stream merge,
    \item channel 22 for flow through an operating fuel cell, and
    \item channel 18 for the concentrations at the outflow.
\end{itemize}
\noindent For simplicity---and without loss of generality for the idle, split, merge and outflow evaluations---we only considered the propagation of hydrogen concentration. Detailed propagation of the other species through a fuel cell channel is provided in \cref{fig:concentration_profiles_oxygen,fig:concentration_profiles_water,fig:concentration_profiles_hydroxide}. Specifically, for channels 6, 5, 11, and 22, the average cross-sectional concentration of the proposed and CFD method were compared along the channel length:
\begin{equation}
    \Delta c(s) = \left| \langle c_\text{p} (s,t)\rangle_t -\langle c_\text{b}(s,t) \rangle_{t}\right|,
\end{equation}
\noindent where \(s\) is the distance along the channel--in flow direction, \(t\) the distance across the channel perpendicular to \(s\) (following right-hand rule), and \(c_p\) and \(c_b\) are the concentration fields of the proposed and baseline methods, respectively. In addition, the \mbox{cross-section} concentration profile at the outlet for both methods is directly compared. Similar to the single-cell case, an operating current density of 0.1~\mbox{A/cm\textsuperscript{2}} was set on all fuel cells. The relative errors between the proposed and CFD methods are illustrated in \cref{fig:PropagationError}, for all three inflow velocities.
 
Overall, obtained results  in \cref{fig:PolarizationCurveResult,fig:concentration_profiles_hydrogen,fig:concentration_profiles_oxygen,fig:concentration_profiles_water,fig:concentration_profiles_hydroxide} clearly show that the proposed method is in line with the CFD simulation. There was little loss of accuracy that could be observed with respect to the CFD simulation. The relative errors in \cref{fig:PropagationError} generally follow a horizontal trend, indicating that little error is added during the propagation. Slight increases in error were observed in the fuel cell channels at higher flow velocities, which can be attributed to the boundary condition approximation described in \cref{sec:BCModels}. Conversely, errors at the outflow decrease with higher velocity, which is consistent with the desired flow sculpting regime, as diffusion time is reduced. Local spikes at the start and end of the channels indicate that errors occur at channel intersections, an observation also described in~\cite{takken2025abstract}. However, as pointed out in~\cite{takken2025abstract}, CFD does not always provide the ground-truth and a similar reduced-order method was more in line with measurements obtained from experiments on a fabricated device than with CFD results.

Furthermore, the runtime improvement for the single cell case is significant, considering that the CFD simulator took 63 seconds to simulate the cell at the operating current density of 0.1 A/cm\textsuperscript{2}, whereas the proposed method only took 0.01 seconds. It should be noted that the CFD simulator likely employed multiple CPU cores in parallel for the simulation, whereas the abstract method only used a single core. Whether the advantage in runtime performance remains for larger stacks is considered next, in \cref{sec:Performance}.

\begin{table*}[t]
    \small
    \centering
    \caption{Runtimes - Newton-Raphson enabled}
    \label{tab:RuntimesSlow}
    \def\arraystretch{1.0}
    \setlength{\tabcolsep}{2pt}
    \makebox[\linewidth]{
        \begin{tabular}{@{\,}r@{\qquad}r@{\qquad}r@{\qquad}r@{\qquad}r@{\qquad}r@{\qquad}r@{\qquad}r@{\qquad}r@{\qquad}r@{\,}}
            & \multicolumn{9}{c}{\textit{\textbf{Runtimes [mm:ss.ss]}}}\\
            \cmidrule{2-10}
            & \multicolumn{3}{c}{$\text{r}_\text{DAG}=0.0$} & \multicolumn{3}{c}{$\text{r}_\text{DAG}=\sqrt{\tilde{n}}$} & \multicolumn{3}{c}{$\text{r}_\text{DAG}=1.0$} \\
            \cmidrule(r{15pt}){2-4} \cmidrule(r{15pt}){5-7} \cmidrule{8-10} 
            \textit{\textbf{Size}} & $\text{r}_\text{Tree}$ = 0.0 & $\text{r}_\text{Tree}$ = 0.5  & $\text{r}_\text{Tree}$ = 1.0 & $\text{r}_\text{Tree}$ = 0.0 & $\text{r}_\text{Tree}$ = 0.5  & $\text{r}_\text{Tree}$ = 1.0 & $\text{r}_\text{Tree}$ = 0.0 & $\text{r}_\text{Tree}$ = 0.5 & $\text{r}_\text{Tree}$ = 1.0  \\
            \midrule
            \begin{filecontents*}{Data/runtimes_slow.csv}
size, resA, resB, resC, resD, resE, resF, resG, resH, resI
4, 00:00.26, 00:00.26, 00:00.25, 00:00.27, 00:00.26, 00:00.26, 00:00.28, 00:00.27, 00:00.27
16, 00:01.01, 00:01.05, 00:01.07, 00:00.09, 00:01.11, 00:01.06, 00:01.07, 00:01.11, 00:01.05
64, 00:04.20, 00:04.30, 00:04.16, 00:04.44, 00:04.48, 00:04.30, 00:04.60, 00:04.34, 00:04.49
256, 00:16.62, 00:17.40, 00:17.21, 00:17.91, 00:17.87, 00:17.93, 00:19.55, 00:18.48, 00:18.14
1024, 02:37.23, 02:00.65, 01:24.57, 04:10.94, 02:21.81, 01:31.39, 03:50.47, 01:41.23, 01:29.85
2048, 21:52.09, 10:00.05, 05:43.77, 20:11.84, 06:17.89, 05:54.12, 18:43.34, 10:53.05, 05:53.39
4096, -,-,32:14.75,-,-,32:28.91,-,-,32:58.59
\end{filecontents*}
\csvreader[head to column names, late after line=\\] 
            {Data/runtimes_slow.csv}{} 
            {\size & \resA & \resB & \resC & \resD & \resE & \resF & \resG & \resH & \resI} 
            \bottomrule
        \end{tabular}
    }
\end{table*}

\begin{table*}[t]
    \small
    \centering
    \caption{Runtimes - Newton-Raphson disabled}
    \label{tab:RuntimesFast}
    \def\arraystretch{1.0}
    \setlength{\tabcolsep}{2pt}
    \makebox[\linewidth]{
        \begin{tabular}{@{\,}r@{\qquad}r@{\qquad}r@{\qquad}r@{\qquad}r@{\qquad}r@{\qquad}r@{\qquad}r@{\qquad}r@{\qquad}r@{\,}}
            & \multicolumn{9}{c}{\textit{\textbf{Runtimes [mm:ss.ss]}}}\\
            \cmidrule{2-10}
            & \multicolumn{3}{c}{$\text{r}_\text{DAG}=0.0$} & \multicolumn{3}{c}{$\text{r}_\text{DAG}=\sqrt{\tilde{n}}$} & \multicolumn{3}{c}{$\text{r}_\text{DAG}=1.0$} \\
            \cmidrule(r{15pt}){2-4} \cmidrule(r{15pt}){5-7} \cmidrule{8-10} 
            \textit{\textbf{Size}} & $\text{r}_\text{Tree}$ = 0.0 & $\text{r}_\text{Tree}$ = 0.5  & $\text{r}_\text{Tree}$ = 1.0 & $\text{r}_\text{Tree}$ = 0.0 & $\text{r}_\text{Tree}$ = 0.5  & $\text{r}_\text{Tree}$ = 1.0 & $\text{r}_\text{Tree}$ = 0.0 & $\text{r}_\text{Tree}$ = 0.5 & $\text{r}_\text{Tree}$ = 1.0  \\
            \midrule
            \begin{filecontents*}{Data/runtimes_fast.csv}
size, resA, resB, resC, resD, resE, resF, resG, resH, resI
4, 00:00.26, 00:00.26, 00:00.26, 00:00.27, 00:00.27, 00:00.26, 00:00.27, 00:00.28, 00:00.29
16, 00:01.05, 00:01.01, 00:01.04, 00:01.11, 00:01.09, 00:01.10, 00:01.15, 00:01.07, 00:01.08
64, 00:04.23, 00:04.11, 00:04.21, 00:04.28, 00:04.21, 00:04.46, 00:04.51, 00:04.54, 00:04.39
256, 00:16.91, 00:16.35, 00:16.55, 00:18.18, 00:17.88, 00:17.95, 00:17.91, 00:18.19, 00:17.17
1024, 01:07.66, 01:05.40, 01:09.16, 01:11.02, 01:11.04, 01:09.75, 01:12.63, 01:09.10, 01:13.04
4096, 04:52.83, 05:26.89, 05:16.85, 05:38.12, 05:00.82, 04:54.13, 05:38.47, 05:30.44, 05:40.82
16384, 19:28.53, 21:23.83, 22:12.37, 22:49.83, 21:32.20, 23:43.20, 22:54.74, 23:09.82, 22:37.10
\end{filecontents*}
\csvreader[head to column names, late after line=\\] 
            {Data/runtimes_fast.csv}{} 
            {\size & \resA & \resB & \resC & \resD & \resE & \resF & \resG & \resH & \resI} 

            \bottomrule
        \end{tabular}
    }
\end{table*}

\section{Scaling performance}
\label{sec:Performance}
To showcase the scaling potential of the proposed method, first, we discuss the different flow network and electrical network configurations that are considered for a stack of \(n\) cells. Afterwards, the runtimes needed to simulate these configurations are provided and the performance is discussed.

\subsection{Considered Stack Configurations}
There are many possible configurations for the fluid and electrical networks, defined in \cref{sec:Stack}, to construct a fuel cell stack. Therefore, to provide a performance evaluation, we considered various network configurations for the fluid network and for the electrical network, where both networks contain the same number of fuel cells~\(n\). More precisely:
\begin{itemize}
    \item The flow network is defined as a \emph{DAG} and every series connection adds a concentration dependency in the system. To this end, we define \(r_\text{DAG}\) as the ratio of graph height to the graph size. That is, a flow network that is completely parallel, and has no concentration dependencies between the fuel cells, has a ratio \(r_\text{DAG}=0.0\). Similarly, a flow network that is fully connected in series has ratio \(r_\text{DAG}=1.0\) and \(n-1\) concentration dependencies. 
    \item The electrical network is defined as a \emph{Tree}, which can have both series and parallel connections between cells. Similar to the flow network, the ratio of parallel and series connections is of interest and we define \(r_\text{Tree}\) as the ratio of series connections in the electrical network.
\end{itemize}
\hfill\\
\noindent For the runtime evaluation, we consider the cases where \(r_\text{DAG}\) can be either 0.0, \(\sqrt{\tilde{n}}\), or 1.0, and \(r_\text{Tree}\) can be either 0.0, 0.5, or 1.0, resulting in nine different combinations of network configurations. Here, \(\tilde{n}\) is chosen such that \(\sqrt{\tilde{n}}\) is the nearest integer number to~\(\sqrt{n}\) and the resulting \emph{DAG} has a pseudo-square shape. The network configurations were generated randomly for a given size and ratio \(r\).

Furthermore, due to the high complexity and dimensionality of the design space, as well as the strong dependence of convergence behavior on the stack operating current, the number of convergence iterations for the scaling procedure was fixed at ten. Accordingly, the simulator evaluates \(n\) cells across nine distinct network configurations, performing ten iterations for each configuration.

\subsection{Runtime Results}
Similar to the simulation of the single cell, the scaled stacks were simulated on a single core to evaluate the method's scalability. The runtimes for each of the nine different network configurations are listed for stack sizes ranging from \(n=4\) to \(n=4,096\) in \cref{tab:RuntimesSlow}. 

The results clearly show the superiority of the proposed abstract simulation method over CFD. While, as covered above, CFD typically requires more than a minute to simulate a \emph{single} cell (and usually can only scale up to a few dozen cells, if at all), the proposed simulation approach can simulate stacks composed of \emph{hundreds} of cells in less than a minute. From \cref{tab:RuntimesSlow}, we observe that the method scales about~\(n^2+n\) for \(r_\text{Tree}=1.0\). This is most likely due to the quadratic scaling of the Newton-Raphson method coming from the jacobian matrix, combined with the propagation operation. Moreover, we observe that the method scales worse with lower \(r_\text{Tree}\), which could be explained by the large number of unknown currents in a system with a large number of parallel connections and, hence, an increasingly larger jacobian matrix in \cref{eq:NewtonRaphson1}.

Considering that the Newton-Raphson method was here solved using a direct solver, and that an iterative solver can solve large sparse systems (which is the case) far more efficiently, we performed a similar scaling experiment for the method, with the Newton-Raphson solving step disabled. This allows us to evaluate the scalability of the proposed abstract method, separate from the implemented method for solving the linear system. The obtained runtimes are listed in \cref{tab:RuntimesFast} and demonstrate that even thousands of cells can be simulated in just a few minutes---stacks with $16,384$ cells in less than an hour, all on a single CPU core. It should be evident that even higher scales could be possible using a parallel implementation of the method. This clearly shows that the proposed method scales well and can, indeed, simulate large scale microfluidic stacks within a reasonable time frame. 




In fact, these results allow us, for the first time, to consider instances of practical value. Recalling the power requirements and corresponding tentative stack size requirements given in Table I, the proposed method would be suitable for the design space exploration of microfluidic fuel cell stacks for UAVs or household applications. Especially, if we consider the runtime improvements for a parallel implementation and if we introduce modular \emph{sub-stacks} that could be designed at a smaller scale (e.g., 1.0\(\times \text{10}^\text{5}\) cells) and connected to provide the required power. While it might be premature to draw conclusions for other applications, the proposed approach significantly advances the state of the art, which previously only allowed for the simulation of stacks composed of a single-digit number of cells. This advancement provides a foundation for further improvements in simulation methods and the development of automated design space exploration methods, potentially enabling the design of stacks for those applications.

\section{Conclusion and Future Work}
\label{sec:Conclusion}
Microfluidic fuel cells present a promising solution for clean and scalable energy production, but their adoption is currently limited by the computational demands of existing simulation tools. In this work, we introduced an abstract simulation approach that enables efficient, \mbox{reduced-order} modeling of large scale fuel cell stacks without compromising accuracy. By simplifying the simulation of individual cells and extending this abstraction to full stacks, the method offers a practical and scalable alternative to traditional \mbox{CFD-based} simulations. Validation against CFD results confirmed its accuracy, and a scalability analysis highlighted its potential for use in \mbox{early-stage} design and optimization. By overcoming the computational limitations of conventional methods, the proposed method represents a significant advancement in the state of the art, allowing, for the first time, efficient simulation of large scale microfluidic fuel cell stacks---including first instances of practical value for macroscopic systems. As microfabrication and integration technologies continue to improve, such simulation methods can play a key role in accelerating the development and deployment of microfluidic fuel cells in \mbox{real-world} energy systems.

The idea and the method presented in this work open up several possibilities for future research. Firstly, the design space is currently only explored manually, and an automated design method for microfluidic fuel cell stack could significantly accelerate the development of prototypical systems. Secondly, the compartmentalized nature of a microfluidic fuel cell stack, could allow for a better flow control inside the stack, and, possibly, lead to more efficient and lighter fuel cell systems. However, a prototypical system would need to be produced, to confirm the theoretical advantages of such a system. The scalable simulation method outlined in this work serves as a solid foundation for this and related lines of research.


\renewcommand{\baselinestretch}{0.96}
{\footnotesize
\bibliographystyle{ieeetr}
{%
\bibliography{references}%
}
}
\end{document}